\documentclass[aps,prb,twocolumn,floatfix,showpacs,superscriptaddress]{revtex4}
\usepackage[dvips]{color}
\usepackage{amsmath,amssymb,amsfonts,bm,graphicx,latexsym}
\usepackage[hypertex,breaklinks=true,colorlinks=false]{hyperref}

\newcommand{\kvec}{\bm{k}}
\newcommand{\Nuc}{N_\mathrm{uc}}
\newcommand{\Rvec}{{\bm{R}}}
\newcommand{\rvec}{{\bm{r}}}
\newcommand{\evec}{{\bm{e}}}
\newcommand{\eff}{\mathrm{eff}}
\newcommand{\hc}{\mathrm{h.c.}}

\newcommand{\ua}{\uparrow}
\newcommand{\da}{\downarrow}

\newcommand{\Svec}{\bm{S}}

\newcommand{\bra}[1]{\langle #1 |}
\newcommand{\ket}[1]{| #1 \rangle}
\newcommand{\bracket}[2]{\langle #1 | #2 \rangle}

\newcommand{\Dcal}{{\mathcal D}}

\newcommand{\Mcal}{{\mathcal M}}

\newcommand{\Ocal}{{\mathcal O}}

\newcommand{\qt}{\tilde{q}}
\newcommand{\dt}{\tilde{d}}

\newcommand{\PRL}[3]{Phys. Rev. Lett. {\bf #1},
\href{http://link.aps.org/abstract/PRL/v#1/e#2}{#2} (#3)}
\newcommand{\PRLp}[3]{Phys. Rev. Lett. {\bf #1},
\href{http://link.aps.org/abstract/PRL/v#1/p#2}{#2} (#3)}

\newcommand{\PRB}[3]{Phys. Rev. B {\bf #1},
\href{http://link.aps.org/abstract/PRB/v#1/e#2}{#2} (#3)}
\newcommand{\PRBp}[3]{Phys. Rev. B {\bf #1},
\href{http://link.aps.org/abstract/PRB/v#1/p#2}{#2} (#3)}
\newcommand{\PRBR}[3]{Phys. Rev. B {\bf #1},
\href{http://link.aps.org/abstract/PRB/v#1/e#2}{#2} (R) (#3)}

\newcommand{\PRE}[3]{Phys. Rev. E {\bf #1},
\href{http://link.aps.org/abstract/PRE/v#1/e#2}{#2} (#3)}

\newcommand{\arXiv}[1]{arXiv:\href{http://arxiv.org/abs/#1}{#1}}

\newcommand{\JPSJ}[3]{J. Phys. Soc. Jpn. {\bf #1},
\href{http://jpsj.ipap.jp/link?JPSJ/#1/#2/}{#2} (#3)}

\begin{document}


\title{
Ferromagnetically-coupled dimers on the 
distorted Shastry-Sutherland lattice: Application to (CuCl)LaNb$_2$O$_7$}

\author{Shunsuke Furukawa}
\affiliation{Department of Physics, University of Toronto, Toronto, Ontario M5S 1A7, Canada}

\author{Tyler Dodds}
\affiliation{Department of Physics, University of Toronto, Toronto, Ontario M5S 1A7, Canada}

\author{Yong Baek Kim}
\affiliation{Department of Physics, University of Toronto, Toronto, Ontario M5S 1A7, Canada}
\affiliation{School of Physics, Korea Institute for Advanced Study, Seoul 130-722, Korea}

\date{\today}



\begin{abstract}
A recent study [Tassel {\it et al.}, Phys. Rev. Lett. {\bf 105}, 167205 (2010)] has proposed a remarkable spin model for (CuCl)LaNb$_2$O$_7$, 
in which dimers are ferromagnetically coupled to each other on the distorted Shastry-Sutherland lattice. 
In this model, the intra-dimer exchange coupling $J>0$ is antiferromagnetic, 
while the inter-dimer exchange couplings are ferromagnetic and take different values, $J_x,J_y<0$, in the two bond directions. 
Anticipating that the highly frustrated character of this model may lead to a wide range of behaviors in (CuCl)LaNb$_2$O$_7$ and related compounds, 
we theoretically investigate the ground state phase diagram of this model in detail using the following three approaches: 
a strong-coupling expansion for small $J_x$ and $J_y$, exact diagonalization for finite clusters, and a Schwinger boson mean field theory. 
When $|J_x|, |J_y|\lesssim J$, the system stays in a dimer singlet phase with a finite spin gap. 
This state is adiabatically connected to the decoupled-dimer limit $J_x=J_y=0$.  
We show that the magnetization process of this phase depends crucially 
on the spatial anisotropy of the inter-dimer couplings.  
The magnetization shows a jump or a smooth increase for weak and strong anisotropy, respectively, 
after the spin gap closes at a certain magnetic field. 
When $|J_x|$ or $|J_y|\gtrsim J$, 
quantum phase transitions to various magnetically ordered phases 
(ferromagnetic, collinear stripe, and spiral) occur. 
The Schwinger boson analysis demonstrates that quantum fluctuations split the classical degeneracy of different spiral ground states. 
Implications for (CuCl)LaNb$_2$O$_7$ and related compounds are discussed 
in light of our theoretical results and existing experimental data.
\end{abstract}
\maketitle


\section{Introduction}\label{sec:introduction}

The search for exotic quantum states in two-dimensional spin systems with frustrated interactions 
has been of great recent interest.\cite{Diep05,Lacroix11} 
Among various materials studied recently, 
$S=\frac12$ layered copper oxyhalides 
(CuX)A$_{n-1}$B$_n$O$_{3n+1}$ 
offer an interesting family of frustrated magnets with rich variety of behaviors. 
In this family, each magnetic CuX layer (with X=Cl,Br) is sandwiched by nonmagnetic layers, 
forming an ideal two-dimensional structure. 
In each layer, the Cu$^{2+}$ ions form a square lattice of $S=\frac12$ spins, 
and a competition between ferromagnetic and antiferromagnetic interactions 
is anticipated from the small Curie-Weiss temperatures (relative to other characteristic energy scales) commonly observed in this family. 
Extensive experimental investigations have uncovered 
a collective singlet ground state with a spin gap in (CuCl)LaNb$_2$O$_7$ (Refs.~\onlinecite{Kageyama05,Kageyama05_mag,Kitada07,Yoshida07}), 
a collinear stripe magnetic order in (CuBr)LaNb$_2$O$_7$ (Ref.~\onlinecite{Oba06}), 
and a magnetization plateau at $1/3$ of the saturated moment 
in (CuBr)Sr$_2$Nb$_3$0$_{10}$ (Ref.~\onlinecite{Tsujimoto07}). 
Furthermore, chemical substitution was used to observe quantum phase transitions between magnetic and non-magnetic ground states.\cite{Uemura09,Kitada09,Tsujimoto10}  
It would be interesting to determine what kind of spin models can capture the wide variety of physics in this family. 
A frustrated $J_1$-$J_2$ model on the square lattice with ferromagnetic $J_1$ and antiferromagnetic $J_2$, 
as initially postulated for this family, 
does not seem to exhibit a spin gap or a magnetization plateau.\cite{Shannon06}

\begin{figure}
\begin{center}
\includegraphics[width=0.40\textwidth]{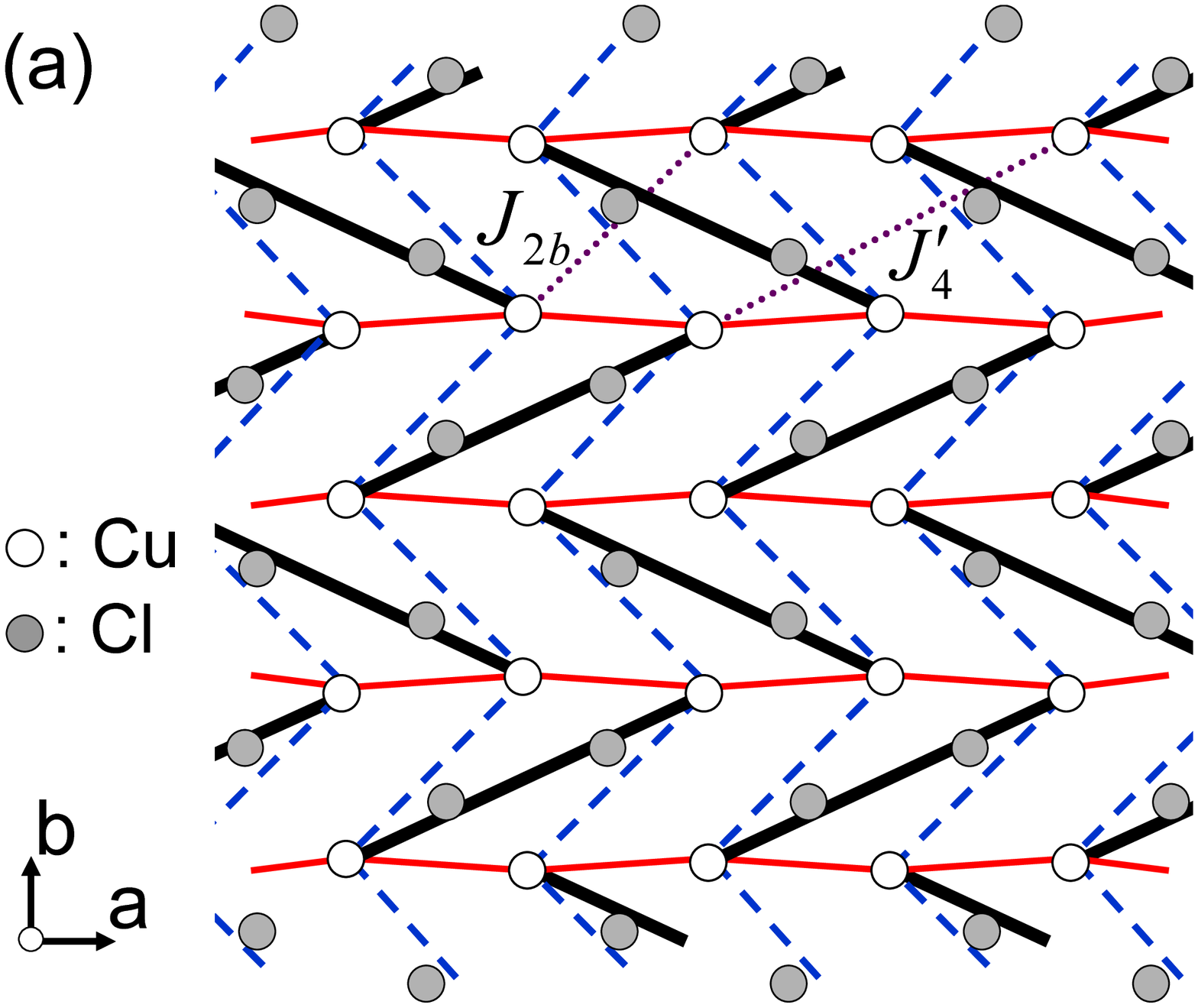}\\
\vspace{2mm}
\includegraphics[width=0.37\textwidth]{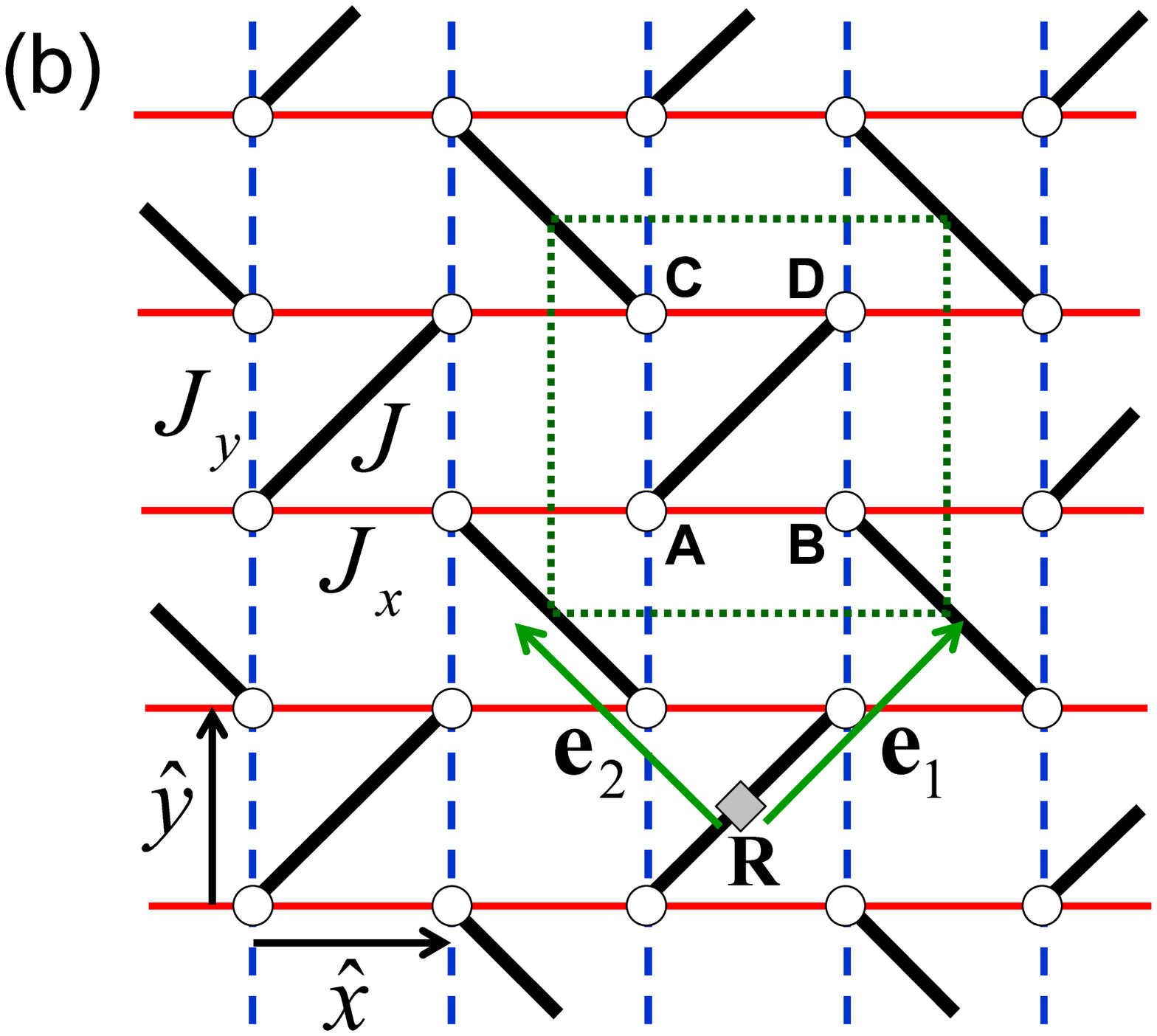}
\end{center}
\caption{(Color online) 
(a) Schematic picture of a single magnetic layer of (CuCl)LaNb$_2$O$_7$. 
Thick solid ($J$), thin solid ($J_x$), and broken ($J_y$) lines indicate 
the three major couplings revealed in the electronic structure calculation of Ref.~\onlinecite{Tassel10}. 
Other interactions $J_{2b}$ and $J_4'$ shown by dotted lines will also be considered in Sec.~\ref{sec:compound}. 
(b) Topologically equivalent Shastry-Sutherland picture of the three major couplings $J$, $J_x$, and $J_y$. 
A dotted green plaquette indicates a unit cell. 
The four sublattices are labeled as $A$, $B$, $C$, and $D$. 
}
\label{fig:lattice}
\end{figure}

A recent study by Tassel {\it et al.}\cite{Tassel10}
has proposed a remarkable microscopic structure in (CuCl)LaNb$_2$O$_7$. 
The x-ray and neutron diffraction studies have identified a considerable distortion of the Cu-Cl bonds;  
consequently, the unit cell is doubled along $a$ and $b$ axes, as shown in Fig.~\ref{fig:lattice}(a). 
Based on the obtained crystal structure, a density functional calculation was carried out 
to construct the microscopic spin model. 
The dominant interaction was then found to be the fourth-neighbor antiferromagnetic exchange coupling $J$, 
which pairs spins into dimer singlets shown as thick lines in Fig.~\ref{fig:lattice}(a). 
The calculation also suggested that the next leading couplings are 
ferromagnetic couplings (denoted by $J_x$ and $J_y$ in this paper) between these dimers, 
which are shown by solid and broken lines in Fig.~\ref{fig:lattice}(a). 
Remarkably, the resultant interaction network has the structure of the distorted Shastry-Sutherland lattice\cite{Shastry81} 
as shown in Fig.~\ref{fig:lattice}(b).\cite{Comment_Tsirlin}



In this paper, 
we study a spin-$\frac12$ model of coupled dimers on the distorted Shastry-Sutherland lattice, 
as the simplest starting point for understanding the interplay of competing interactions in (CuCl)LaNb$_2$O$_7$ and related compounds. 
We consider the Heisenberg model in a magnetic field, 
\begin{equation}\label{eq:H}
 H=\sum_{(i,j)} J_{ij} \Svec_i \cdot \Svec_j - h\sum_i S^z_i,
\end{equation}
where $(i,j)$ runs over all the bonds in Fig.~\ref{fig:lattice}(b), 
and $J_{ij}=J$, $J_x$, and $J_y$ for diagonal, horizontal and vertical bonds, respectively. 
We are primarily concerned with the case of antiferromagnetic $J>0$ and ferromagnetic $J_x,J_y<0$;  
however, results for other signs of $J_x$ and $J_y$ are presented alongside for comparison.
A common viewpoint for dealing with coupled dimer systems is 
to regard the triplet excitation at each dimer as a particle (called a ``triplon'') 
and to describe the system as a Bose gas of such particles.\cite{Giamarchi08,Rice02}  
A notable feature of the Shastry-Sutherland lattice is 
the strong suppression of the triplon hopping due to frustration.\cite{Miyahara99,Miyahara03}  
In the antiferromagnetic case with $J>0$ and $J_x=J_y>0$, 
the localized nature of the triplons gives rise to various fractional plateaux in the magnetization process,\cite{Miyahara00,Momoi00,Misguich01}  
which are experimentally observed\cite{Kageyama99} in SrCu$_2$(BO$_3$)$_2$. 
Frustration also exist in the case of our interest, with ferromagnetic $J_x,J_y<0$.  
The consequence of frustration in the ferromagnetic case has not been addressed in previous studies. 


Our main results for the ferromagnetic case $J_x,J_y<0$ are as follows. 
When $|J_x|, |J_y|\lesssim J$, the system stays in a dimer singlet phase with a finite spin gap. 
This state is adiabatically connected to the decoupled dimer limit $J_x=J_y=0$. 
The magnetization process of this phase depends crucially  
on the spatial anisotropy of the inter-dimer couplings: 
the magnetization shows a jump and a smooth increase for weak and strong anisotropy, respectively, 
after the spin gap closes at a certain magnetic field. 
When $|J_x|$ or $|J_y|\gtrsim J$, 
the spin gap of the dimer singlet phase closes, 
and quantum phase transitions to various magnetically ordered phases 
(ferromagnetic, collinear stripe, and spiral) occur. 
It is demonstrated that quantum fluctuations split the classical degeneracy of different spiral ground states. 
The detailed phase diagrams of the classical and quantum models are constructed.
These results are based on a strong-coupling expansion for weak $J_x$ and $J_y$, 
exact diagonalization for small clusters, 
and a Schwinger boson mean field theory. 
Using the existing data on the magnetization process and the triplon bandwidth in (CuCl)LaNb$_2$O$_7$, 
and comparing with the corresponding theoretical results, 
we provide a consistency check for the appropriateness of the proposed spin model \eqref{eq:H}.


The rest of the paper is organized as follows. 
In Sec.~\ref{sec:phase}, we present the classical and quantum phase diagrams of the model \eqref{eq:H} 
and summarize the main results of the paper. 
In Sec.~\ref{sec:classical}, we discuss the details of the computations for the classical model and
the resulting ground states.
In Sec.~\ref{sec:SC}, we perform the strong-coupling expansion of the quantum model to derive 
an effective low-energy Hamiltonian. 
This effective model will be very useful in determining the global phase diagram of the quantum model 
and in estimating relevant physical quantities. 
In Sec.~\ref{sec:ED}, the numerical analysis of the quantum model based on the exact diagonalization
is presented. Some parts of the phase boundaries are determined quite accurately within this approach. 
In Sec.~\ref{sec:SchwingerMFT}, the Schwinger boson mean field theory analysis of the model
is reported. 
These three techniques are collectively used to understand the quantum phase diagram. 
In Sec.~\ref{sec:compound}, we compare these theoretical results on the quantum model with experimental data on (CuCl)LaNb$_2$O$_7$. 
We conclude the paper in Sec.~\ref{sec:discussion}. 
Appendix~\ref{app:exactGS} explains an exact solution available in the isotropic case $J_x=J_y$.  
In Appendix~\ref{app:Hartree}, the Hartree variational approach is discussed. 
In Appendix~\ref{app:saturation}, the saturation field in the magnetization process 
is determined exactly by considering a single-magnon excitation from the fully polarized state.  

\section{Phase diagrams and main results} \label{sec:phase}

\begin{figure}
\begin{center}
\includegraphics[width=0.40\textwidth]{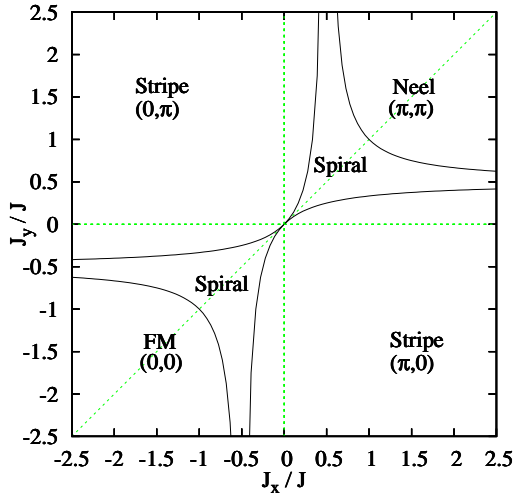}
\end{center}
\caption{(Color online) 
Classical phase diagram of the distorted Shastry-Sutherland model \eqref{eq:H} with $h=0$. 
Green dashed lines at $J_x=0$, $J_y=0$, $J_x=J_y$ are guides for eyes. 
}
\label{fig:phase_cl}
\end{figure}

In this section we present the classical and quantum phase diagrams of the model \eqref{eq:H} 
in Figs.~\ref{fig:phase_cl}, \ref{fig:phase_ED}, and \ref{fig:phase_SB}, 
and summarize the main results of the paper. 
We assume $J>0$ and $h\ge 0$ throughout the paper. 
With the assumption of $J>0$, frustration at the classical level 
occurs only when $J_xJ_y>0$. 
When we discuss the isotropic case $J_x=J_y$, 
we denote these two parameters by $J'$. 

\subsection{Classical phase diagram}\label{subsec:classical}

The classical spin model is defined by replacing all the spin operators in Eq.~\eqref{eq:H} 
by classical $O(3)$ vectors of length $S$. 
Assuming $h=0$ for simplicity, 
the ground state phase diagram in Fig.~\ref{fig:phase_cl} is determined exactly for different signs of $J_x$ and $J_y$. 
The details of our analysis will be presented in Sec.~\ref{sec:classical}. 
Below we summarize the main characteristics of each phase. 

\paragraph*{Collinear stripe phases.} 
These phases appear naturally in the unfrustrated regions $J_xJ_y<0$ 
and penetrate into some parts of the frustrated regions $J_xJ_y>0$. 
The spin configuration has a propagation vector of $(k_x,k_y):=(\kvec\cdot\hat{x},\kvec\cdot\hat{y})=(0,\pi)$ or $(\pi,0)$, 
where $\hat{x}$ and $\hat{y}$ are defined in Fig.~\ref{fig:lattice}(b). 
In the ground state with $(0,\pi)$, for example, 
the system forms stripes of up or down spins running along the $x$ direction, 
while up and down alternate in the $y$ direction. 

\paragraph*{N\'eel phase.} 
This phase appears for large positive $J_x/J$ and $J_y/J$, 
where there is a minimal effect from the $J$ coupling. 
The spin configuration is collinear and has a propagation vector of $(k_x,k_y)=(\pi,\pi)$, 
as usual for the square lattice antiferromagnet. 

\paragraph*{Ferromagnetic (FM) phase.} 
This phase appears for large negative $J_x/J$ and $J_y/J$, 
where ferromagnetic $J_x$ and $J_y$ dominate over antiferromagnetic $J$. 
The spins are all aligned in the same direction and have a propagation vector of $(k_x,k_y)=(0,0)$. 

\paragraph*{Spiral phases.} 
These phases appear when the magnitudes of $J$, $J_x$, and $J_y$ are comparable.  
This is the case where the effect of the frustration is most prominent. 
The ground state is a coplanar spiral state with an incommensurate propagation vector,  
as originally found in the isotropic case $J_x=J_y$ in Ref.~\onlinecite{Shastry81}.   
A notable feature of the spiral phase is that 
two kinds of spiral ground states are degenerate, 
apart from the trivial degeneracy associated with the global $O(3)$ rotation of spins. 
One ground state is the ``$x$-spiral'' in Fig.~\ref{fig:spiral_Qx}(c), 
where spins rotate by a uniform angle $Q_x$ in the $x$ direction and alternating angles $\pm Q_y$ in the $y$ direction. 
The other is the ``$y$-spiral'', which is defined by interchanging the roles of $x$ and $y$ directions. 
Remarkably, in the anisotropic case $J_x\ne J_y$,  
the $x$- and $y$-spirals are not related to each other by any symmetry operation. 
Their degeneracy therefore comes from the particular geometry of the lattice. 

We note that the classical phase boundaries in Fig.~\ref{fig:phase_cl} 
are symmetric with respect to the sign flips $J_{x,y} \to -J_{x,y}$. 
This is because the signs of $J_x$ and $J_y$ can both be flipped 
by reversing the spins on the $A$ and $D$ sites of all units cells,
as seen in Fig.\ref{fig:lattice}(b). 
This transformation is {\it not} allowed in the quantum case, 
since the simultaneous reversal of all the spin components $(S^x,S^y,S^z)\to (-S^x,-S^y,-S^z)$
changes the commutation relations of spin operators. 
The quantum phase diagram presented next therefore depends on the signs of $J_x$ and $J_y$. 

\subsection{Quantum phase diagram}

\begin{figure}
\begin{center}
\includegraphics[width=0.50\textwidth]{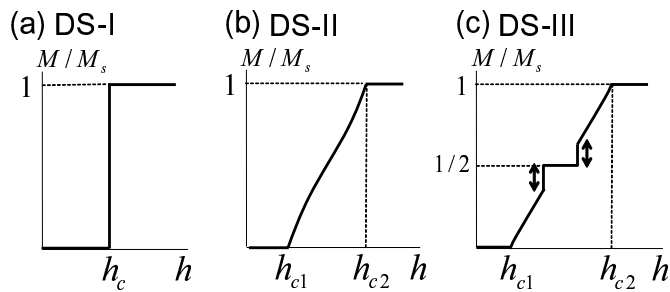}
\end{center}
\caption{
Sketches of magnetization processes in the dimer singlet phase:  
(a) a jump, (b) a smooth increase, and (c) a plateau at $1/2$ of the saturated moment $M_s$.
}
\label{fig:mag_h}
\end{figure}

The quantum model with $S=\frac12$ has been studied intensively 
in the isotropic antiferromagnetic case $J_x=J_y(\equiv J')$. 
A remarkable feature of this model is that for $h=0$ and $0\le J'/J \le 1/2$, 
the ground state of the model is exactly given by the product of dimer singlets\cite{Shastry81}
\begin{equation}\label{eq:DS}
 \ket{ \Psi_{\rm DS} } = \prod_\Rvec \ket{s}_\Rvec,
\end{equation}
where $\Rvec$ labels a dimer and $\ket{s}_\Rvec$ is the singlet state on the dimer. 
This solution can be extended to the ferromagnetic region $-1<J'/J<0$, as described in Appendix~\ref{app:exactGS}. 
A notable difference between the ferromagnetic- and antiferromagnetic-$J'$ cases occurs 
when a magnetic field $h$ is applied.
In the antiferromagnetic-$J'$ case, the magnetization process shows 
various fractional plateaux,\cite{Miyahara00,Momoi00,Misguich01}  
which can be viewed as density wave formations of triplon excitations. 
In the ferromagnetic-$J'$ case, 
the magnetization process shows a jump from the dimer singlet state \eqref{eq:DS} to the fully saturated state, as in Fig.~\ref{fig:mag_h}(a). 
As will be discussed in Sec.~\ref{subsec:SC2}, this can be viewed as a consequence of the phase separation of triplons. 

\begin{figure}
\begin{center}
\includegraphics[width=0.40\textwidth]{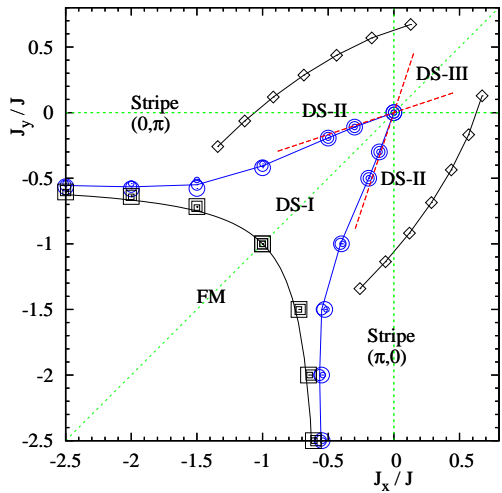}
\end{center}
\caption{(Color online) 
Phase diagram of the distorted Shastry-Sutherland model~\eqref{eq:H} in the spin-$\frac12$ case, 
determined by the strong coupling expansion (Sec.~\ref{sec:SC}) and exact diagonalization (Sec.~\ref{sec:ED}) analyses. 
Green dashed lines at $J_x=0$, $J_y=0$, $J_x=J_y$ are guides for the eye. 
The dimer singlet (DS) phase is divided into three regions, I, II, and III, 
which are characterized by the magnetization processes in Fig.~\ref{fig:mag_h}. 
Red broken lines around the origin indicate the region boundaries 
determined from the the first-order effective Hamiltonian (see Fig.~\ref{fig:phase_SC1}). 
The square and circular symbols are based on exact diagonalization for the number of spins, $N_s=16,20,6\times4$. 
The three different symbol sizes are in the order of $N_s$. 
The classical ferromagnetic phase boundary is superposed on the square symbols, showing good agreement. 
The diamond symbols are the boundaries between the DS-II regions and the stripe phases, 
determined by the fidelity susceptibility analysis in Fig.~\ref{fig:fid}. 
Narrow spiral phases may appear between the DS-I region and the stripe phases for large $|J_x|/J$ or $|J_y|/J$. 
For this reason, the DS-stripe phase boundaries (diamond symbols) are not calculated beyond $J_x/J$ or $J_y/J \approx -1.3$. 
}
\label{fig:phase_ED}
\end{figure}

\begin{figure}
\begin{center}
\includegraphics[scale=0.85]{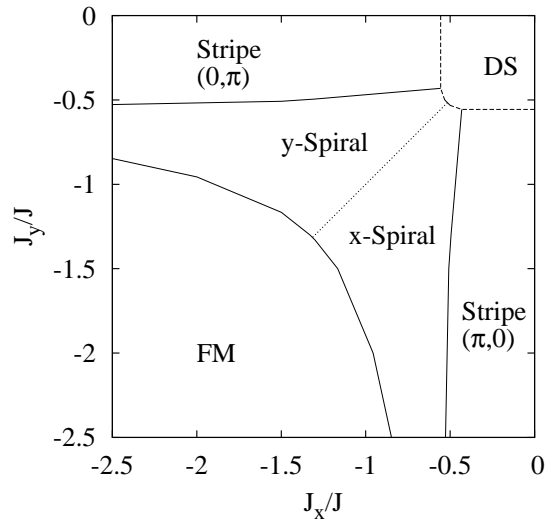}
\end{center}
\caption{
Phase diagram of the distorted Shastry-Sutherland model~\eqref{eq:H}, 
determined by a Schwinger boson mean field theory (Sec.~\ref{sec:SchwingerMFT}) for $S=0.5$. 
Solid lines indicate second-order transitions, while broken lines indicate first-order transitions. 
The isolated dimer singlet (DS) state is found for small $|J_x|,|J_y|\lesssim 0.5J$, 
while the ferromagnetic state stabilizes for larger $|J_x|$ and $|J_y|$. 
The stripe state expands from the classical case to fill $|J_x|$ or $|J_y| \lesssim 0.55 J$. 
The spiral state exists in the middle. 
The $x$-spiral appearing for $|J_x|<|J_y|$ has an incommensurate long-range order in the $x$ direction as in Fig.~\ref{fig:spiral_Qx}.  
The $y$-spiral similarly appears for $|J_x|>|J_y|$. 
The dotted line separates the two spiral phases. 
}
\label{fig:phase_SB}
\end{figure}

Our major interest is in the ferromagnetic case of $J_x,J_y<0$ with spatially anisotropic couplings. 
For comparison, we will also include some results for other signs of $J_x$ and $J_y$. 
The ground state phase diagrams are presented in Figs.~\ref{fig:phase_ED} and \ref{fig:phase_SB}. 
Figure~\ref{fig:phase_ED} is based on 
a strong-coupling expansion for weak $J_x$ and $J_y$ (Sec.~\ref{sec:SC}) 
and exact diagonalization for finite clusters (Sec.~\ref{sec:ED}). 
When $|J_x|, |J_y|\lesssim J$,  
the system stays in a dimer singlet phase with a finite spin gap, 
which is adiabatically connected to the decoupled-dimer limit $J_x=J_y=0$. 
This phase is split into three regions (I, II, and III), 
in terms of the behaviors of the magnetization processes, as shown in Fig.~\ref{fig:mag_h}. 
In the DS-I region, the magnetization $M$ shows a jump as in Fig.~\ref{fig:mag_h}(a). 
In the DS-II region, the magnetization $M$ smoothly increases towards saturation 
after the spin gap closes at a certain magnetic field, as in Fig.~\ref{fig:mag_h}(b). 
Finally, in the DS-III region, the magnetization process shows a plateau at $1/2$ of the saturated moment $M_s$
(and more plateaux can appear near the isotropic case\cite{Miyahara00,Momoi00,Misguich01} $J_x=J_y$). 
When $|J_x|/J$ or $|J_y|/J\gtrsim1$,  
the spin gap of the dimer singlet phase closes, leading to various magnetically ordered states. 
The characteristics of the ferromagnetic and collinear stripe phases are 
rather similar to those in the classical model. 
The transition lines from the dimer singlet phase to these two phases 
are determined by exact diagonalization with relatively good accuracy 
(square and diamond symbols in Fig.~\ref{fig:phase_ED}).

The exact diagonalization for small systems, however, is not adequate
to find incommensurate spiral phases that may intervene 
between the stripe, dimer singlet, and ferromagnetic phases. 
In order to get insight into this question,
we performed a Schwinger boson mean-field analysis (Sec.~\ref{sec:SchwingerMFT}). 
In this approach, the spin magnitude $S$ can be varied freely, 
and one can discuss how the classical phase diagram in the limit $S\to\infty$ (Fig.~\ref{fig:phase_cl})  
changes as quantum fluctuations are gradually taken into account. 
The phase diagram for $S=0.5$ is shown in Fig.~\ref{fig:phase_SB}. 
In this result, we find that the spiral phases do appear in some parts of the phase diagram, 
while the presence of various magnetically ordered phases are somewhat
exaggerated, as expected in mean-field methods.
Providing an interesting difference from the classical case, 
quantum fluctuations split the classical degeneracy of the two spiral states, 
favoring the $x$- or $y$-spirals for $|J_x|<|J_y|$ and $|J_x|>|J_y|$, respectively. 
Although the mean-field method is not useful in discussing the precise locations of the phase boundaries, 
it is natural to expect that these spirals appear in narrow regions between the DS-I region and the stripe phases in Fig.~\ref{fig:phase_ED}, 
particularly for large $|J_x|/J$ or $|J_y|/J$.  


\subsection{Relation to the experiments on (CuCl)LaNb$_2$O$_7$}

In experiments on (CuCl)LaNb$_2$O$_7$, 
the magnetization increases smoothly beyond a critical magnetic field until it reaches the saturated value.\cite{Kageyama05_mag} 
This suggests that, at least within the $J$-$J_x$-$J_y$ model, 
the ground state should belong to the DS-II region as described in Fig.~\ref{fig:phase_ED} 
and there should be substantial anisotropy in $J_x$ and $J_y$. 
Considering the magnetization data\cite{Kageyama05_mag} and the measured energy range $\Delta \epsilon$ of the triplon excitations,\cite{Tassel10} 
and comparing them with theoretical results, 
we find that within the $J$-$J_x$-$J_y$ model,  
one of $J_x$ and $J_y$ may be ferromagnetic while the other may be antiferromagnetic. 
However, depending on how to interpret the existing experimental data, 
the energy range of the triplon excitations can be larger; 
if this is the case, $J_x$ and $J_y$ can be both ferromagnetic. 
We plot the triplon dispersions for the cases of 
(a) ferromagnetic $J_x(<0)$ and antiferromagnetic $J_y(>0)$ 
and (b) both ferromagnetic $J_x,J_y(<0)$ (Fig.~\ref{fig:band_JJxJy}). 
These plots can be used to test the $J$-$J_x$-$J_y$ model further and to determine the signs of $J_x$ and $J_y$,  
when more detailed information of the triplon excitations is provided from experiments. 
We also discuss the effects of other exchange couplings $J_{2b}$ and $J_4'$ shown in Fig.~\ref{fig:lattice}(a), 
which are also contained in the model of Ref.~\onlinecite{Tassel10}. 
These couplings induce oscillating behaviors in the triplon dispersions, 
which can be used as fingerprints of their existence. 
The details of these analyses are presented in Sec.~\ref{sec:compound}.


\section{Classical ground state} \label{sec:classical}

\begin{figure}
\begin{center}
\includegraphics[width=0.50\textwidth]{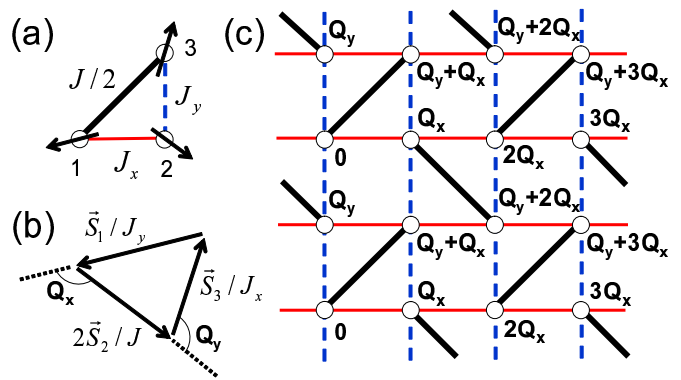}
\end{center}
\caption{(Color online) 
(a) A triangular part of the Hamiltonian. 
(b) Sketch of how to solve the three-spin problem. 
A coplanar ground state is obtained by forming a triangle using three spins. 
(c) A classical spiral ground state with a pitch angle $Q_x$ in the $x$ direction (``$x$-spiral''). 
The number at each site indicates the angle (in the $xy$ plane) of the spin, 
where $Q_x$ and $Q_y$ are given by Eq.~\eqref{eq:QxQy}. 
}
\label{fig:spiral_Qx}
\end{figure}

In this section we describe the exact solution for the classical ground state. 
The phase diagram is presented in Fig.~\ref{fig:phase_cl}, 
with the phases summarized in Sec.~\ref{subsec:classical}. 
For simplicity, we restrict ourselves to the case of $h=0$ and $J, J_x,J_y>0$. 
The case of $J>0$ and $J_x,J_y<0$ can be treated in parallel 
by applying the spin reversal transformation on the $A$ and $D$ sites,
as explained in the last paragraph of Sec.~\ref{subsec:classical}. 
In the case of $J_xJ_y<0$, the system is not frustrated 
and the ground states are naturally determined as collinear stripe states. 

To find the ground state of the classical model with $J,J_x,J_y>0$, 
we decompose the Hamiltonian into triangular parts 
and solve a single-triangle problem shown in Fig.~\ref{fig:spiral_Qx}(a). 
Here the diagonal coupling $J$ is divided by two, since it is shared by two neighboring triangles. 
By rewriting the three-spin energy $E_\triangle$ as 
\begin{equation}
 E_\triangle = \frac{J_x J_y J}4 \left( \frac{\Svec_1}{J_y} + \frac{2\Svec_2}{J} + \frac{\Svec_3}{J_x} \right)^2 +{\rm const.}, 
\end{equation} 
we find that the ground state of $E_\triangle$ is obtained 
by minimizing the length of the vector 
$\Svec_1/J_y + 2\Svec_2/J + \Svec_3/J_x$. 
When the vector lengths $2S/J$, $S/J_x$, and $S/J_y$ have comparable magnitudes, 
the three vectors can form a triangle so that the above sum vanishes, as shown in Fig.~\ref{fig:spiral_Qx}(b). 
This leads to a coplanar spin configuration as shown in Fig.~\ref{fig:spiral_Qx}(a), 
where the spins rotate counterclockwise by $Q_x$ and then by $Q_y$ when moving between sites $1\to 2\to 3$. 
Another configuration where the spins instead rotate clockwise by the same angles also gives a ground state. 
Here $Q_x$ and $Q_y$ satisfy the relations 
\begin{subequations}
\begin{gather}
 \frac1{J_y}\sin Q_x = \frac1{J_x} \sin Q_y, \\
 \frac1{J_y}\cos Q_x + \frac1{J_x} \cos Q_y = -\frac2J, 
\end{gather}
\end{subequations}
which are solved as 
\begin{subequations}\label{eq:QxQy}
\begin{gather}
 \cos Q_x = - \frac{J_y}J + \frac{J}{4J_x} \left( \frac{J_y}{J_x} - \frac{J_x}{J_y} \right),\\
 \cos Q_y = - \frac{J_x}J + \frac{J}{4J_y} \left( \frac{J_x}{J_y} - \frac{J_y}{J_x} \right). 
\end{gather}
\end{subequations}
Using a ground state of $E_\triangle$ locally on every triangle, 
one can construct a ground state of the whole lattice as in Fig.~\ref{fig:spiral_Qx}(c). 
In this state, the spins rotate by a uniform angle $Q_x$ in the $x$ direction, 
and by alternating angles $\pm Q_y$ in the $y$ direction. 
We call this state the ``$x$-spiral.'' 
Similarly, one can construct the ``$y$-spiral'', 
where the spins rotate by a uniform angle $Q_y$ in the $y$ direction 
and by alternating angles $\pm Q_x$ in the $x$ direction. 
As explained in Sec.~\ref{sec:phase}, 
the degeneracy of these states does not result from symmetry, 
and thus it should be regarded as a consequence of the particular geometry of the lattice. 

When $2/J>1/J_x+1/J_y$ (with $J,J_x,J_y>0$), the three spins of $E_\triangle$ can no longer form a triangle as in Fig.~\ref{fig:spiral_Qx}(b). 
Instead, $\Svec_1$ and $\Svec_3$ align antiparallel to $\Svec_2$, 
forming a collinear configuration with $Q_x=Q_y=\pi$. 
This leads to a N\'eel ground state of the whole lattice. 
Similarly, for $1/J_x>2/J+1/J_y$ and $1/J_y>2/J+1/J_x$, 
one obtains collinear stripe ground states with propagation vectors $(0,\pi)$ and $(\pi,0)$, respectively. 

The phase diagram obtained above for the antiferromagnetic quadrant $J_x,J_y>0$ 
can be mapped onto the ferromagnetic quadrant $J_x,J_y<0$ 
by applying the spin reversal 
mentioned in the last paragraph of Sec.~\ref{subsec:classical}. 
The N\'eel ground state maps onto the ferromagnetic state. 
The $(0,\pi)$ stripe state maps onto the $(\pi,0)$ stripe state. 
A spiral state with $Q_x,Q_y > \pi /2$ maps to a spiral state with $Q_x,Q_y < \pi /2$.
These arguments complete the phase diagram for all possible signs of $J_x$ and $J_y$, 
as shown in Fig.~\ref{fig:phase_cl}. 

\section{Strong-coupling expansion} \label{sec:SC}

In this section we analyze the spin-$\frac12$ model \eqref{eq:H} 
by means of a strong coupling expansion.\cite{Momoi00,Totsuka98,Kolezhuk99,Mila11}.  
In this approach, we start from the limit $J_x=J_y=0$, where dimers are decoupled from each other. 
We then perturbatively include the effects of $J_x$ and $J_y$ 
and derive an effective Hamiltonian of the model \eqref{eq:H}.  
In Sec.~\ref{subsec:Heff}. 
we obtain the effective Hamiltonian up to second order in $J_x$ and $J_y$. 
Then, in Secs.~\ref{subsec:SC1} and \ref{subsec:SC2}, 
we analyze the first- and second-order effective Hamiltonians, respectively, 
to deduce the physical properties of the original Hamiltonian $H$. 
The first-order Hamiltonian is exactly equivalent to the XXZ model on the square lattice. 
Using the known results on the XXZ model, we map out a qualitative phase diagram of the model \eqref{eq:H}. 
The second-order Hamiltonian provides more accurate estimations of physical quantities and phase boundaries than the first-order one. 
The obtained phase boundaries show a remarkable agreement with the exact diagonalization result of Sec.~\ref{sec:ED}. 
The spin gap and the triplon band width derived in Sec.~\ref{subsec:SC2} 
will be used to fit the experimental data of (CuCl)LaNb$_2$O$_7$ in Sec.~\ref{sec:compound}.  

\subsection{Effective Hamiltonian} \label{subsec:Heff}

\begin{figure}
\begin{center}
\includegraphics[width=0.50\textwidth]{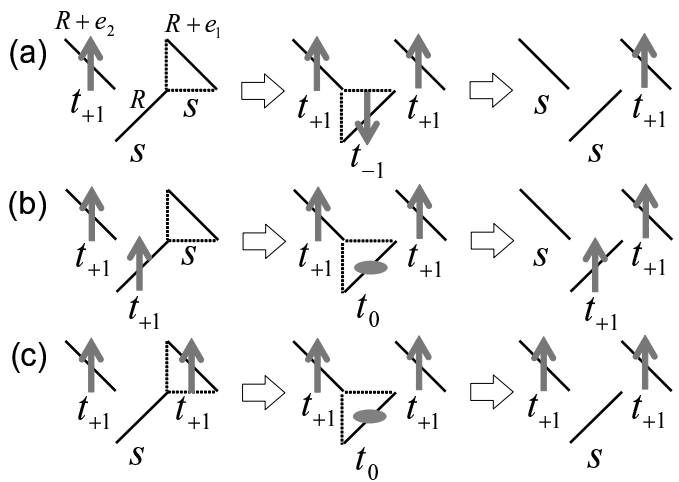}
\end{center}
\caption{
Examples of virtual processes in the second-order perturbation theory, 
which lead to correlated hopping terms (a) $(1-n_R) b_{R+e1}^\dagger b_{R+e2}$ and (b) $n_R b_{R+e1}^\dagger b_{R+e2}$, 
and a three-body interaction term (c) $(1-n_R) n_{R+e1}  n_{R+e2}$. 
In each process, $H_1$ acts first on $R$ and $R+e_1$, as indicated by the dotted lines, to yield the excited state. 
It then acts on $R$ and $R+e_2$ to yield the final state of the process.
}
\label{fig:virtual}
\end{figure}


We start from the limit $J_x=J_y=0$, where dimers are decoupled from each other. 
We derive an effective Hamiltonian of the model \eqref{eq:H} by perturbatively including the effects of $J_x$ and $J_y$. 
We label each dimer by its center position $\Rvec$ as in Fig.~\ref{fig:lattice}(b); 
the two spins on the dimer are then denoted by $\Svec_{\Rvec 1}$ and $\Svec_{\Rvec 2}$. 
When $h=0$, the eigenstates of each dimer consist of a singlet $\ket{s}$ and a triplet $\{ \ket{t_\mu} \}$, 
where $\mu=0,\pm 1$ represents the $z$ component of the total spin, $S_{R1}^z+S_{R2}^z$.  
When a magnetic field $h>0$ is applied, the degeneracy of the triplet is split, 
with $\ket{t_{+1}}=\ket{\!\ua\ua}$ having the lowest energy of the three. 
Then it is useful to focus on the low-energy sector of the Hilbert space 
which consists of $\ket{s}$ and $\ket{t_{+1}}$. 
We apply degenerate perturbation theory to derive an effective Hamiltonian 
in this restricted Hilbert space. 
Specifically, we adjust the magnetic field to $h=J$ 
so that the singlet $\ket{s}$ and the up-polarized state $\ket{t_{+1}}$ become degenerate. 
The unperturbed Hamiltonian is 
\begin{equation}\label{eq:H0}
 H_0 = \sum_\Rvec \big[ J \Svec_{\Rvec 1} \cdot \Svec_{\Rvec2} - J (S^z_{\Rvec1}+S^z_{\Rvec2}) \big]. 
\end{equation}
The remaining part of the Hamiltonian is 
\begin{equation}\label{eq:H1}
 H_1 = \sum_i \big[ J_x \Svec_i \cdot \Svec_{i+\hat{x}} + J_y \Svec_i \cdot \Svec_{i+\hat{y}} - (h-J) S^z_i \big], 
\end{equation}
where $i$ runs over all the sites. 
We treat $H_1$ as a perturbation, assuming that the coefficient of each term is sufficiently smaller than $J$.  
Note that the coefficient of the Zeeman term in $H_1$ is $h-J$ ({\it not} $h$) 
since the magnetic field of magnitude $J$ is included in $H_0$; 
therefore, the perturbation theory is most accurate around the intermediate magnetic field $h=J$. 
The ground state of $H_0$ is macroscopically degenerate with a degeneracy of $2^{2\Nuc}$, 
where $\Nuc$ is the number of unit cells in the system. 
Let $V_0$ be the subspace of the Hilbert space spanned by these states. 
Regarding $\ket{t_{+1}}$ and $\ket{s}$ on each dimer $\Rvec$ as the presence and the vacancy of a particle (``triplon''),  
we treat the system as a hard-core boson gas on the square lattice of dimer centers. 
On each dimer $\Rvec$, 
we define a creation operator $b^\dagger_\Rvec=\ket{t_{+1}}\bra{s}$ and a number operator $n_\Rvec=\ket{t_{+1}}\bra{t_{+1}}$.  
Degenerate perturbation theory up to second order yields an effective Hamiltonian 
\begin{equation}\label{eq:Heff}
 H^\eff = E_0 + H^\eff_1 + H^\eff_2, 
\end{equation} 
where $E_0 =-\frac32 \Nuc J$ is the ground state energy of $H_0$, 
and $H^\eff_1$ and $H^\eff_2$ are the first- and second-order contributions, respectively. 

The first-order contribution $H^\eff_1$ is calculated 
by projecting $H_1$ onto the ground-state manifold $V_0$ of $H_0$: 
\begin{align}\label{eq:Heff1_ori}
 H^\eff_1 =& 
 \sum_{\langle \Rvec\Rvec' \rangle} \left[ -\frac{J_-}2 (b_\Rvec^\dagger b_{\Rvec'} +\hc) + \frac{J_+}2 n_\Rvec n_{\Rvec'} \right] \notag\\
 &- (h-J) \sum_\Rvec n_\Rvec,
\end{align}
where $\langle \Rvec\Rvec' \rangle$ runs over all the nearest-neighbor pairs of dimers, and  
\begin{equation}\label{eq:Jpm}
 J_\pm := \frac{J_x\pm J_y}2. 
\end{equation}
The $J_x$ and $J_y$ couplings produce the triplon hopping and interaction terms between nearest-neighbor dimers. 
The two couplings contribute with opposite signs to the hopping term, 
so the coefficient is proportional to $J_-$.  
This leads to a suppression of the boson hopping near the isotropic case $J_x=J_y$, 
which is a remarkable feature of the Shastry-Sutherland lattice.\cite{Miyahara99,Momoi00}  
The interaction term is repulsive or attractive depending on the sign of $J_+$. 
The Zeeman term $h-J$ in $H_1$ plays the role of the chemical potential of the hard-core boson gas. 
Since this term is diagonal in the particle number basis $\{n_\Rvec\}$, it contributes only to the first-order Hamiltonian $H^\eff_1$. 

To calculate the second-order contribution $H^\eff_2$, 
we take into account various processes 
where a state in $V_0$ is virtually promoted to an excited state of $H_0$ by the operation of $H_1$,
and then comes back to $V_0$ by $H_1$ again. 
Such virtual processes give rise to correlated hopping terms as well as three-body interaction terms, 
as illustrated in Fig.~\ref{fig:virtual}. 
The resulting form of $H^\eff_2$ is 
\begin{widetext}
\begin{align}
 H^\eff_2 = 
 &-\frac{3\Nuc J_-^2}{2J} + \sum_{\langle RR' \rangle} \left[ -\frac{J_+J_-}{2J} (b_R^\dagger b_{R'} +\hc) + \left(\frac{J_+^2}{2J}+\frac{J_-^2}{8J}\right) n_R n_{R'} \right] 
 -\left( \frac{J_+^2}{J} - \frac{J_-^2}{2J} \right) \sum_R n_R \notag\\
 &+ \frac{J_+J_-}{4J} \sum_R \sum_{\nu=1}^4 \left[ 2(-1)^\nu (1-n_R) n_{R+e_\nu} n_{R+e_{\nu+1}}
  + n_{R+e_\nu} (b_R^\dagger b_{R-e_\nu} +\hc) \right] \notag\\
 &+ \sum_{R\in \Dcal_1} \bigg[ \frac{J_+^2}{2J} n_{R+e1}(1-n_R) n_{R-e1} + \frac{J_-^2}{2J} n_{R+e2}(1-n_R) n_{R-e2} 
    + \frac{J_-^2}{8J} (3n_R-1) (b_{R+e1}^\dagger b_{R-e1} + \hc) \notag\\
&~~~+ \left( \frac{J_+^2}{4J} n_R -\frac{J_-^2}{8J} (1-n_R) \right) (b_{R+e2}^\dagger b_{R-e2} + \hc) 
    + \sum_{\nu=1}^4 \left( (-1)^\nu \frac{J_+J_-}{4J} n_R -\frac{J_-^2}{8J} (1-n_R) \right) (b_{R+e_\nu}^\dagger b_{R+e_{\nu+1}} +\hc) \notag\\
&~~~- \frac{J_+^2}{4J} (n_{R+e1}-n_{R-e1}) (b_R^\dagger b_{R+e2} - b_R^\dagger b_{R-e2} +\hc)
    - \frac{J_-^2}{4J} (n_{R+e2}-n_{R-e2}) (b_R^\dagger b_{R+e1} - b_R^\dagger b_{R-e1} +\hc) \bigg] \notag\\
 &+ \sum_{R\in \Dcal_2} \left[ e_1 \leftrightarrow e_2, ~J_-\leftrightarrow -J_- \right].
\end{align}
\end{widetext}
Here, the vectors $\evec_1$ and $\evec_2$ are defined as in Fig.~\ref{fig:lattice}, 
and $\evec_{3,4,5}$ are defined via the relations $\evec_1=-\evec_3=\evec_5$ and $\evec_2=-\evec_4$. 
$\Dcal_1$ and $\Dcal_2$ refer to the sets of dimers oriented in the directions of $\evec_1$ and $\evec_2$, respectively. 
By setting $J_x=J_y$, the effective Hamiltonian presented here coincides with the one derived by Momoi and Totsuka.\cite{Momoi00}


\subsection{First order: qualitative phase diagram} \label{subsec:SC1}

By identifying the hard-core boson operators 
with the spin-$\frac12$ operators via $b_\Rvec^\dagger=s^+_\Rvec$ and $n_\Rvec=s^z_\Rvec+1/2$, 
the first-order effective Hamiltonian $H^\eff_1$ is equivalent to the XXZ model on the square lattice,  
\begin{equation}\label{eq:Heff1}
 H^\eff_1 = \sum_{\langle \Rvec\Rvec' \rangle} \big[ K_{xy} (s^x_\Rvec s^x_{\Rvec'} + s^y_\Rvec s^y_{\Rvec'}) + K_z s^z_\Rvec s^z_{\Rvec'} \big]
 - \tilde{h} \sum_\Rvec s^z_\Rvec,
\end{equation}
with 
\begin{equation}
 K_{xy} = -J_-,~~ K_z = \frac{J_+}2,~~\tilde{h}=h-J-J_+. 
\end{equation}
Note that this model has a higher symmetry than the original model; 
the unit cell has been reduced from two dimers to a single dimer. 
The phase diagram of this model has been studied in detail in the literature.\cite{Matsuda70,Kohno97,Yunoki02} 
By changing $\tilde{h}$ from large negative to large positive values, 
the ground state of $H^\eff_1$ changes between the fully down-polarized ($\langle s_\Rvec \rangle=-1/2$) 
and fully up-polarized ($\langle s_\Rvec \rangle=+1/2$) states. 
In the original model \eqref{eq:H}, this corresponds 
to a change between the dimer singlet state \eqref{eq:DS} with magnetization $M=0$ and the fully polarized state with $M=M_s$. 
The process of this change can be smooth or sudden; 
the details depend on the value of the XXZ anisotropy $K_z/|K_{xy}|$ of $H^\eff_1$, 
and are classified into three cases (I)-(III), described below. 
We note that the sign of $K_{xy}$ is not crucial in the present argument, 
since it can be flipped by applying a gauge transformation to $H^\eff_1$ 
or by exchanging the roles of $J_x$ and $J_y$ in $H$. 

(I) Ferromagnetic Ising case $K_z/|K_{xy}|<-1$.  
At $\tilde{h}=0$, the fully up- and down-polarized states are degenerate in the ground state of $H^\eff_1$. 
Therefore, the magnetization of the effective XXZ model 
shows a jump from $\langle s_\Rvec \rangle=-1/2$ to $+1/2$ 
as $\tilde{h}$ changes its sign from negative to positive. 
This corresponds, in the original model, to a jump of the magnetization $M$ at an intermediate magnetic field $h_c=J+J_+$ as in Fig.~\ref{fig:mag_h}(a). 

(II) XY case $-1< K_z/|K_{xy}| < 1$.  
The ground state of $H^\eff_1$ at $\tilde{h}=0$ is given by a ferromagnetic or antiferromagnetic state in the $xy$ plane 
for $K_{xy}<0$ and $K_{xy}>0$, respectively. 
This XY ordered state changes to a canted state by a magnetic field $\tilde{h}\ne 0$ 
and the magnetization $\langle s_\Rvec \rangle$ varies smoothly between $-1/2$ and $+1/2$ 
in the range $-2(|K_{xy}|+K_z)< \tilde{h} <2(|K_{xy}|+K_z)$. 
Therefore, in the original model, 
the magnetization $M$ increases smoothly from zero to the saturation 
in the field range between $h_{c1}=J-2|J_-|$ and $h_{c2}=J+2(J_++|J_-|)$, as in Fig.~\ref{fig:mag_h}(b). 
The canted ferromagnetic and antiferromagnetic orders in the effective model 
give rise to transverse magnetizations in the original model. 
Specifically, in the field range $h_{c1}<h<h_{c2}$, 
the transverse component shows a stripe order 
with a propagation vector of $(k_x,k_y)=(0,\pi)$ and $(\pi,0)$, for $J_x<J_y$ and $J_x>J_y$, respectively. 
The connection between the magnetic orders of the effective and original models 
is understood most easily from a variational ground state approach, which is presented in Appendix \ref{app:Hartree}. 

(III) Antiferromagnetic Ising case $K_z/|K_{xy}|>1$.   
In contrast to the XY case above, the ground state of $H^\eff_1$ at $\tilde{h}=0$ is given by an antiferromagnetic state polarized
along the $z$ axis, featuring gapped excitations. 
When $\tilde{h}$ is increased above zero, the magnetization $\langle s_\Rvec \rangle$ stays zero 
up to certain critical $\tilde{h}>0$ and then shows a jump in the transition to  
the canted antiferromagnetic state (known as the spin-flopping process).\cite{Matsuda70,Kohno97,Yunoki02} 
A similar thing occurs when decreasing $\tilde{h}$ below zero. 
Therefore, in the original model, the magnetization process shows a plateau at $1/2$ of the saturation, 
accompanied by jumps at the edges, as in Fig.~\ref{fig:mag_h}(c). 
The expressions of $h_{c1}$ and $h_{c2}$ are the same as the XY case (II) above. 

\begin{figure}
\begin{center}
\includegraphics[width=0.50\textwidth]{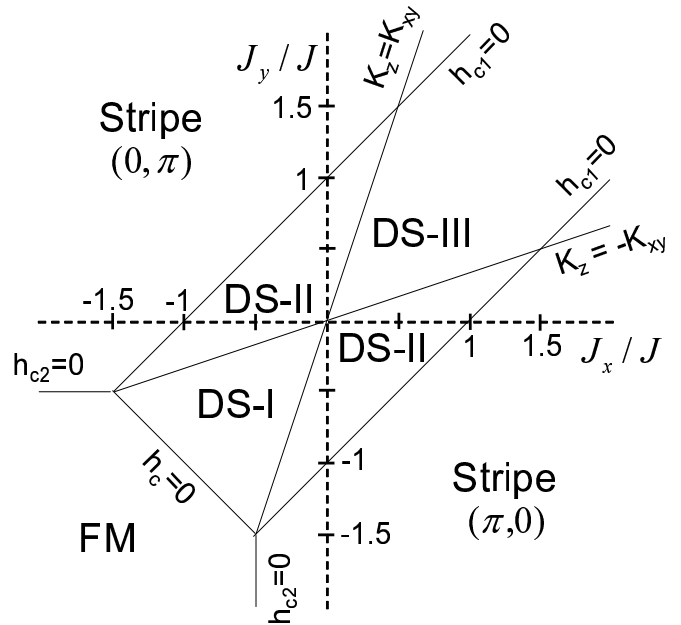}
\end{center}
\caption{
Qualitative phase diagram of the spin-$\frac12$ model \eqref{eq:H}, derived from the first-order effective Hamiltonian $H^\eff_1$
of the strong-coupling expansion. 
In the DS-I and DS-II regions, the magnetization shows a jump and a smooth increase as in Figs.~\ref{fig:mag_h}(a) and (b), respectively. 
In the DS-III region, at least one magnetization plateau is expected to appear at $1/2$ of the saturated moment as in Fig.~\ref{fig:mag_h}(c). 
More plateaux may be expected near the isotropic case.\cite{Miyahara00,Momoi00,Misguich01} 
}
\label{fig:phase_SC1}
\end{figure}

The present argument allows us to map out a qualitative phase diagram as in Fig.~\ref{fig:phase_SC1}. 
The dimer singlet phase appearing for $|J_x|,|J_y|\lesssim J$ 
are divided into three regions (I, II, and III) in terms of the behaviors of the magnetization processes shown in Fig.~\ref{fig:mag_h}. 
The boundaries between the regions are given by $K_z=\pm K_{xy}$ from the above argument. 
In principle, the effective Hamiltonian $H^\eff_1$ can give reliable results for $H$ only when $J_x$ and $J_y$ are sufficiently smaller than $J$. 
Beyond this region, we nonetheless use $H^\eff_1$ to investigate possible instabilities of the dimer singlet state. 
By increasing $|J_x|$ and $|J_y|$ in the DS-I region, the critical field $h_c=J+J_+$ goes to zero, 
indicating a transition to the ferromagnetic phase. 
Similarly, by increasing $|J_x|$ or $|J_y|$ in the DS-II regions, $h_{c1}=J-2|J_-|$ goes to zero, 
leading to the collinear stripe phases. 
The boundaries between the ferromagnetic and stripe phases are determined by the condition $h_{c2}=J+2(J_++|J_-|)=0$. 
The obtained phase diagram in Fig.~\ref{fig:phase_SC1} qualitatively agrees with Fig.~\ref{fig:phase_ED} obtained by exact diagonalization. 
However, some aspects of the phase diagram are not captured in this approach. 
For example, the spiral and N\'eel phases 
(expected from the classical analysis of Sec.~\ref{sec:classical} 
and the Schwinger boson analyses of Refs.~\onlinecite{Albrecht96,Chung01} and Sec.~\ref{sec:SchwingerMFT}) 
do not appear in this approach. 
These phases are beyond the scope of the present perturbative analysis. 


In passing, we note that the phase diagram in Fig.~\ref{fig:phase_SC1} can also be obtained using a Hartree variational state approach,  
which is presented in Appendix \ref{app:Hartree}. 

\subsection{Second order: estimations of physical quantities} \label{subsec:SC2}

We now move to the analysis of the second-order effective Hamiltonian $H^\eff$. On one hand, this
allows us to make more accurate estimations of physical quantities, such as the spin gap, compared to the first-order
case above. 
On the other hand, the new terms appearing in $H^\eff$ may open up possibilities of new phenomena. 
In the isotropic antiferromagnetic case $J_x=J_y>0$, Momoi and Totsuka\cite{Momoi00} 
have shown that correlated hopping processes $n_R b_{R'}^\dagger b_{R''}$ 
induce the formation of bound states of two triplons. 
In the ferromagnetic case of our main interest, however, 
we will argue that the formation of bound states is rather unlikely, 
and that the second-order terms in $H^\eff$ do not change the essential physical properties of the system. 
Instead, the improved accuracy in the estimation of physical quantities 
allows us to make more quantitative comparison with the exact diagonalization results of Sec.~\ref{sec:ED}. 

\begin{figure}
\begin{center}
\includegraphics[width=0.45\textwidth]{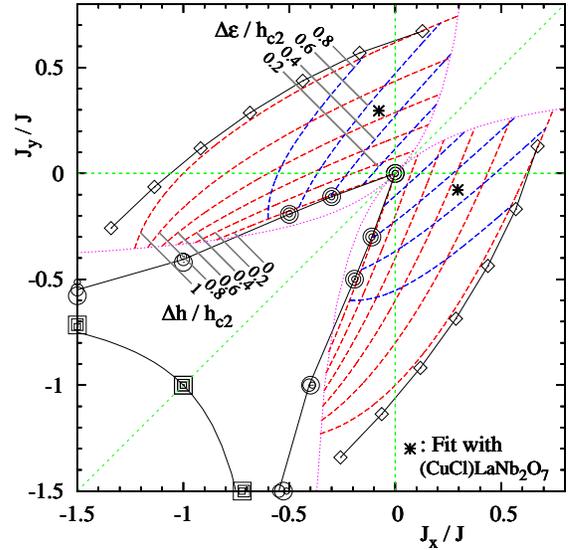}
\end{center}
\caption{(Color online) 
Contour maps of $\Delta h/h_{c2}$ and $\Delta\epsilon/h_{c2}$ in the DS-II regions, 
where the magnetization process shows a smooth increase as in Fig.~\ref{fig:mag_h}(b). 
Here $\Delta h:=h_{c2}-h_{c1}$ is the width between the starting and ending of the magnetization process, 
and $\Delta\epsilon$ is the width of the triplon dispersion. 
The contours are shown only inside the pink dotted lines, where Eq.~\eqref{eq:hc2_stripe} is valid. 
The lines of $\Delta h/h_{c2}=0$ and $1$ correspond to transition lines 
to the DS-I region and the stripe phases, respectively. 
The transition lines obtained from exact diagonalization analysis of Sec.~\ref{sec:ED} are also shown 
(by the same symbols as in Fig.~\ref{fig:phase_ED}). 
Fit with the experimental data of (CuCl)LaNb$_2$O$_7$ yields estimates of $J_x/J$ and $J_y/J$, 
as indicated by the asterisks (see Sec.~\ref{sec:compound}). 
}
\label{fig:hc1}
\end{figure}

We first calculate the energy of a single-triplon excitation. 
By restricting to the single-triplon sector, 
we can neglect all the terms of the forms, $n_Rn_{R'}$, $n_R(1-n_{R'})n_{R''}$ and $n_R b_{R'}^\dagger b_{R''}$. 
The resulting Hamiltonian contains only the chemical potential $\mu$ 
and the nearest-neighbor, second-neighbor, and third-neighbor hoppings, 
whose amplitudes are given respectively by $t$, $-2t'$, and $-t'$.  
These parameters are given by 
\begin{equation}
 \mu = h-J+\frac{J_+^2}J - \frac{J_-^2}{2J},~~
 t = -\frac{J_-}2 \left( 1 + \frac{J_+}{J} \right),~~
 t' = \frac{J_-^2}{8J}. 
\end{equation}
The unit cell is again reduced to a single dimer; however, this occurs only in a single-triplon sector. 
This Hamiltonian leads to a triplon dispersion, 
\begin{equation}
\begin{split}
 \epsilon_{\bm{k}} = & -\mu + 2t (\cos k_1 + \cos k_2) \\
 & - 4t' [ \cos (k_1+k_2) + \cos (k_1-k_2)] \\
 & - 2t' (\cos 2k_1 + \cos 2k_2),
\end{split}
\end{equation}
with $k_\nu := \bm{k}\cdot \evec_\nu~(\nu=1,2)$. 
Using $k_x:=\bm{k}\cdot\hat{x}$ and $k_y:=\bm{k}\cdot\hat{y}$, this is rewritten as 
\begin{equation}\label{eq:band_JJxJy}
 \epsilon_{\bm{k}} = -\mu + 4t' + 4t\cos k_x\cos k_y - 16t' \cos^2 k_x \cos^2 k_y. 
\end{equation}
This dispersion has the minimum energy
\begin{equation}
 \epsilon_{\rm min} = -\mu-4|t|-12t', 
\end{equation}
and the band width 
\begin{equation}
 \Delta \epsilon = 8|t| = 4|J_-| \left( 1+\frac{J_+}J \right)
\end{equation}
(if $-J_+ + 2|J_-|<J$). 
The triplon condensation field $h_{c1}$ in the DS-II regions is determined by the condition $\epsilon_{\rm min}=0$, 
leading to 
\begin{equation}
 h_{c1} = J - 2|J_-| - \frac{(J_+ + |J_-|)^2}J. 
\end{equation}
The saturation field $h_{c2}$ is calculated exactly in Appendix~\ref{app:saturation}. 
Using $h_{c2}=J+2J_++2|J_-|$ in Eq.~\eqref{eq:hc2_stripe}, 
which is valid in the collinear stripe regions of the classical model, 
the width $\Delta h$ between $h_{c1}$ and $h_{c2}$ is determined as 
\begin{equation}
 \Delta h := h_{c2}-h_{c1} = 2J_+ + 4|J_-| + \frac{(J_+ + |J_-|)^2}J. 
\end{equation}
In Fig.~\ref{fig:hc1}, 
we display the contour maps of $\Delta h/h_{c2}$ and $\Delta\epsilon/h_{c2}$ calculated in this manner. 
Since $\Delta h=\Delta\epsilon=0$ in the decoupled-dimer limit $J_x=J_y=0$, 
the two parameters indicate to what extent the system is separated from this limit. 
However, the two parameters show slightly different behaviors. 
The contours of $\Delta h/h_{c2}$ and $\Delta\epsilon/h_{c2}$ are approximately parallel 
to the lines $J_x=3J_y$ (or $3J_x=J_y$) and $J_x=J_y$, respectively. 
The lines of $\Delta h/h_{c2}=0$ corresponds to the transition between the DS-I and DS-II regions. 
The lines of $\Delta h/h_{c2}=1$ corresponds to the quantum phase transitions into magnetically ordered states 
associated with the triplon condensation. 
These boundaries agree well with the exact diagonalization results plotted together in Fig.~\ref{fig:hc1}.  
This means that our perturbative calculation likely gives a good estimation of $h_{c1}$ 
up to rather large values of $|J_x|/J$ and $|J_y|/J$. 

Next, we discuss the role of the correlated hopping processes $n_R b_{R'}^\dagger b_{R''}$, 
which appear in the second-order effective Hamiltonian. 
In the isotropic antiferromagnetic case $J_x=J_y>0$, Momoi and Totsuka\cite{Momoi00} 
have shown that these processes induce the formation of bound states of two triplons. 
Namely, two triplons pair together and achieve a lower energy than two independent triplons. 
In this case, as the magnetic field is increased, 
the bound states condense before the triplons do, 
leading to a bond-nematic order.\cite{Shannon06,Ueda07} 
It is interesting to look for this possibility in the ferromagnetic case of our interest. 
We performed exact diagonalization analysis of the original spin model \eqref{eq:H} 
on various points in the ferromagnetic case $J_x,J_y<0$, 
and searched for the existence of bound states. 
However, we found no indication of the formation of bound states. 
This result can be interpreted from the study of Schmidt {\it et al}..\cite{Schmidt06} 
These authors have analyzed a simple hard-core boson model on the square lattice 
containing a single-particle hopping $t$, a correlated hopping $\tilde{t}$, and the nearest-neighbor repulsion $V$. 
They have demonstrated numerically that the correlated hopping process 
enhances the tendency towards phase separation as well as boson pairing; 
a jump of the magnetization [Fig.~\ref{fig:mag_h}(c)] in the DS-I region 
indeed occurs as a result of the phase separation.\cite{Comment_PS} 
It was shown\cite{Schmidt06} that the correlated hopping process induces the boson paring 
only when a modest magnitude of $V>0$ is introduced.
In the ferromagnetic region $J_x,J_y<0$ of the model \eqref{eq:H}, 
the nearest-neighbor interaction is attractive ($V<0$), 
and hence phase separation and a standard superfluid state are more likely to occur than the boson pair-condensed state.

\section{Exact diagonalization} \label{sec:ED}

\begin{figure}
\begin{center}
\includegraphics[width=0.50\textwidth]{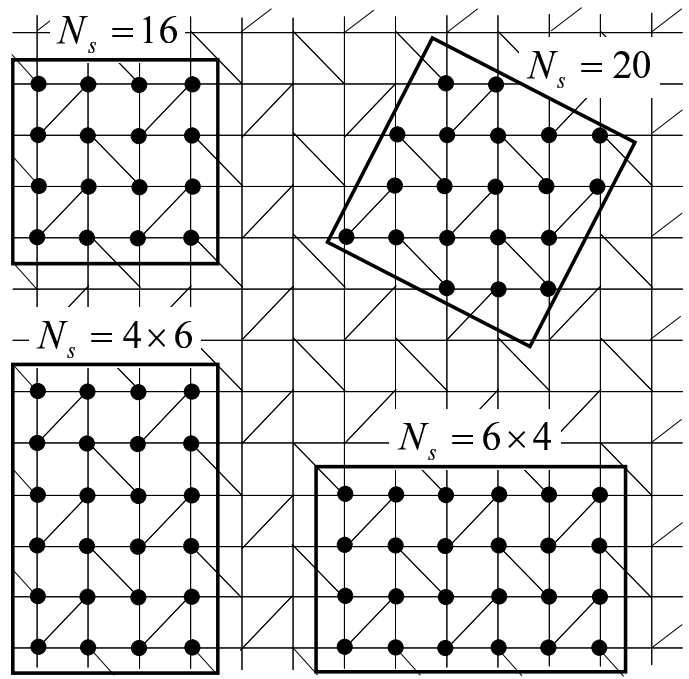}
\end{center}
\caption{
Clusters with $N_s=16,20,4\times6$, and $6\times4$ used for exact diagonalization analyses. 
}
\label{fig:clusters}
\end{figure}

\begin{figure}
\begin{center}
\includegraphics[width=0.40\textwidth]{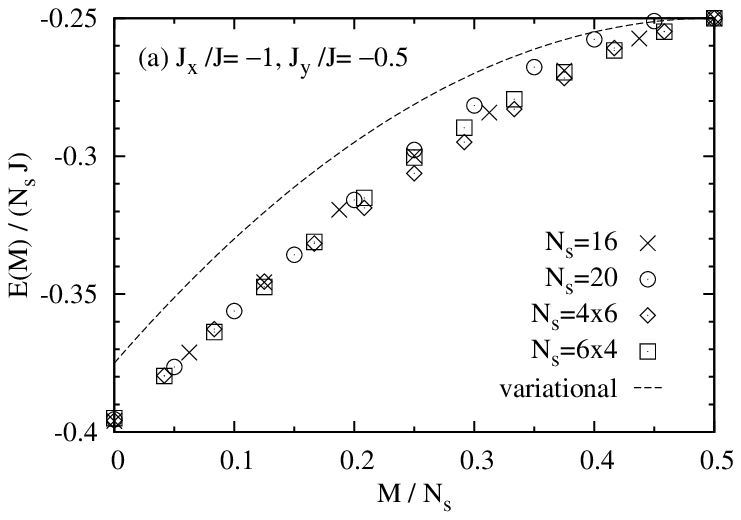}
\includegraphics[width=0.40\textwidth]{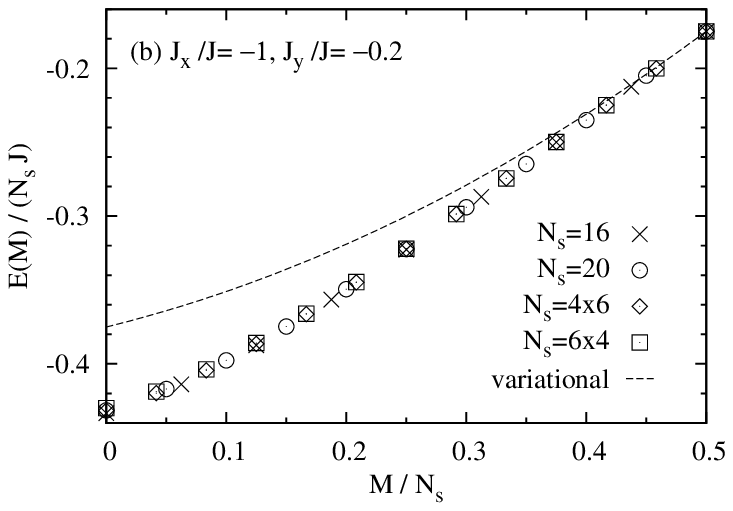}
\end{center}
\caption{
Ground state energy $E(M)$ as a function of the magnetization $M$  
for (a) $J_x=-1$ and $J_y=-0.5$ and (b) $J_x=-1$ and $J_y=-0.2$. 
A concave curve in (a) leads to a jump in the magnetization as in Fig.~\ref{fig:mag_h}(a), 
while a convex curve in (b) leads to a smooth increase as in Fig.~\ref{fig:mag_h}(b). 
Broken lines indicate the variational estimates using the ansatz \eqref{eq:Hartree}. 
}
\label{fig:ene_M}
\end{figure}

\begin{figure}
\begin{center}
\includegraphics[width=0.40\textwidth]{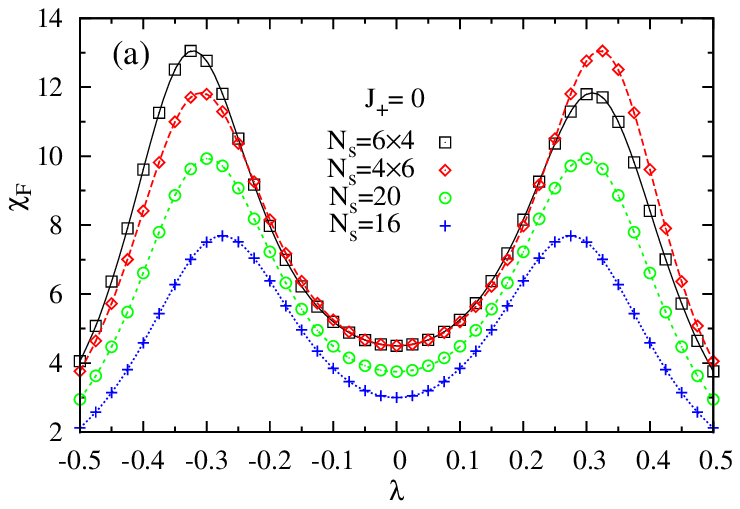}
\includegraphics[width=0.40\textwidth]{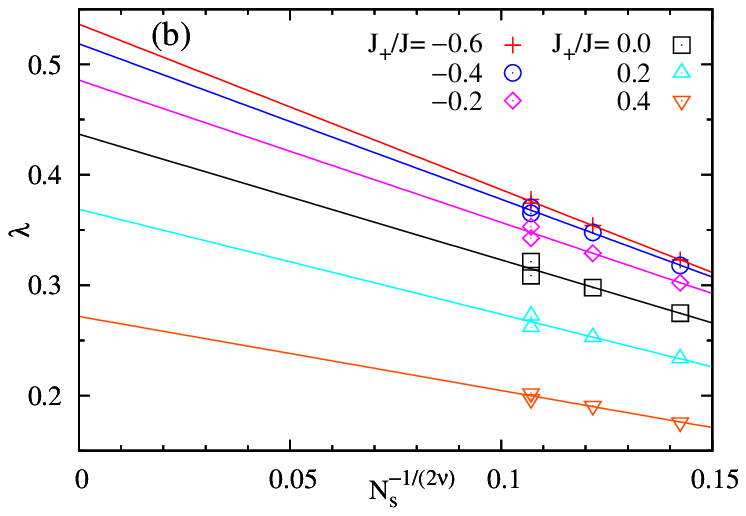}
\end{center}
\caption{(Color online) 
(a) Fidelity susceptibility $\chi_F$ versus $\lambda:=J_-/J$ for $J_+=0$. 
(b) Extrapolation of the peak position of $\chi_F$ into $N_s\to\infty$ using Eq.~\eqref{eq:fid_scale}. 
}
\label{fig:fid}
\end{figure}

The strong coupling expansion in the previous section has provided 
controlled analyses for small $|J_x|/J$ and $|J_y|/J$. 
To investigate the ground-state phase diagram of the model in a wide parameter space, 
we have performed exact (Lancz\"os) diagonalization calculations in finite-size clusters. 
The shapes of the clusters are shown in Fig.~\ref{fig:clusters}, 
where $N_s$ is the total number of spins in the system. 
In every cluster, periodic boundary conditions are imposed at the edges. 
The clusters with $N_s=16$ and $N_s=20$ are invariant under simultaneous operations of 
$90^\circ$ lattice rotation and interchange of $J_x$ and $J_y$. 
Therefore, these clusters treat $J_x$ and $J_y$ equally, and are less biased in studying the ground-state properties
than the clusters with $N_s=24$, which do not have this invariance. 
However, to investigate the size dependence of the calculated quantities, 
we analyze the $N_s=24$ clusters as well.

\subsection{Magnetization process} \label{subsec:ED_mag}

We first investigate the magnetization process in the dimer singlet phase. 
The magnetization process is determined by calculating the ground-state energy $E(M)$
as a function of the total magnetization $M:=\sum_i S^z_i$ as in Fig.~\ref{fig:ene_M}. 
To determine the magnetization $M$ for a given field $h$ from this figure, we first find
the line of slope $h/J$ that touches the curve in such a way as to minimize the
vertical axis intercept. 
The magnetization is then found at the touching point of the line and the curve. 
A concave curve, as in Fig.~\ref{fig:ene_M}(a), leads to a jump of $M$ as in Fig.~\ref{fig:mag_h}(a), with 
\begin{equation}
 h_c= \frac{E(N_s/2)-E(0)}{N_s/2}. 
\end{equation}
A convex curve, as in Fig.~\ref{fig:ene_M}(b), leads instead to a smooth increase of $M$ as in Fig.~\ref{fig:mag_h}(b) with 
\begin{equation}
 h_{c1} = E(1) - E(0), ~~~
 h_{c2} = E(N_s/2) - E(N_s/2-1). 
\end{equation}
A convenient way to judge whether the curve is concave or convex 
is to examine which of $h_c$ and $h_{c1}$ defined above is smaller. 
By plotting $h_c$ and $h_{c1}$ as functions of model parameters ($J_x/J$ or $J_y/J$) and finding the crossing point, 
one can determine the boundary between the DS-I and DS-II regions, 
as indicated by circular symbols in Fig.~\ref{fig:phase_ED}. 
Here, the data for $N_s=16$, $20$, and $6\times4$ are shown by symbols of different sizes in order of $N_s$. 
The obtained boundaries show little dependence on the system size $N_s$. 
In Fig.~\ref{fig:ene_M}, we also plot the estimate of the energy using the Hartree variational ansatz \eqref{eq:Hartree}.  
We find that this ansatz can capture the qualitative shapes of the curves. 


As $J_x$ or $J_y$ increases, 
the spin gap of the system becomes smaller and eventually vanishes, 
leading to magnetically ordered states. 
The transition points to the ferromagnetic state can be determined accurately 
by finding the level crossing between the singlet ground state and the fully polarized state. 
These transition points are plotted by square symbols in Fig.~\ref{fig:phase_ED}. 
The obtained boundary agrees very well with the classical ferromagnetic phase boundary.\cite{comment_FM}
The transitions to the collinear stripe states cannot be detected by a level crossing, 
and we instead calculate the fidelity susceptibility, as presented below.

\subsection{Transition to the stripe phases} \label{subsec:ED_stripe}

The fidelity susceptibility\cite{You07,Albuquerque10} 
shows a peak at the transition point,
which we use to detect the phase boundary. 
Since the phase boundaries are parallel to the line $J_x=J_y$ in the qualitative phase diagram of Fig.~\ref{fig:phase_SC1},  
it is efficient to probe the parameter space in the direction perpendicular to this line. 
Therefore we fix $J_+/J$ and change the parameter $\lambda:=J_-/J$, 
where $J_\pm:=(J_x\pm J_y)/2$ as defined in Eq.~\eqref{eq:Jpm}. 
The fidelity between two parameter points $\lambda$ and $\lambda'$ is defined as an overlap between the ground states, 
\begin{equation}
 F(\lambda,\lambda') = |\bracket{\Psi(\lambda)}{\Psi(\lambda')}|. 
\end{equation}
When the two points are close enough, 
this quantity has an expansion 
\begin{equation}
 F(\lambda,\lambda+\delta\lambda) = 1-\frac{(\delta\lambda)^2}2 \chi_F + \dots,
\end{equation}
which allows us to define the fidelity susceptibility\cite{You07}  
\begin{equation}
 \chi_F (\lambda) = -2 \lim_{\delta\lambda \to 0} \frac{\ln F(\lambda,\lambda+\delta\lambda)}{\delta\lambda^2}. 
\end{equation}
The result for $J_+=0$ is presented in Fig.~\ref{fig:fid}(a).
We observe a peak in $\chi_F (\lambda)$, which indicates a phase transition; 
this peak shifts gradually and grows sharper with increasing $N_s$. 
The peak position $\lambda_c(N_s)$ in a finite-size system is expected to approach the transition point $\lambda_c$
with increasing $N_s$. 
In Fig.~\ref{fig:fid}(b), we extrapolate the data of $\lambda_c(N_s)$ 
into $N_s\to\infty$, assuming the scaling form\cite{Albuquerque10}  
\begin{equation}\label{eq:fid_scale}
 [\lambda_c-\lambda_c(N_s)] \sim N_s^{-1/(2\nu)}, 
\end{equation}
with $\nu\approx 0.711$ 
for an $O(3)$ quantum phase transition, corresponding to triplon condensation. 

The obtained stripe phase transition points are plotted by diamond symbols in Fig.~\ref{fig:phase_ED}. 
The phase boundary crosses the vertical axis at $(J_x/J,J_y/J) \approx (0,0.63)$. 
When $J_x=0$ and $J_y>0$, the present model is equivalent to a Heisenberg antiferromagnet on the honeycomb lattice 
with spatially anisotropic exchange couplings. 
Quantum Monte Carlo analysis of such a model\cite{Jiang09} has given a precise estimate $J_y/J\approx 0.576$ for the transition point. 
Our result roughly agrees with this, although it slightly overestimates the range of the dimer singlet phase. 

The classical phase diagram in Fig.~\ref{fig:phase_cl} suggests that 
there can be a narrow spiral phase between the ferromagnetic and collinear stripe phases. 
Since incommensurate spin correlations existing in the spiral phase 
are difficult to treat in the small system sizes available to exact diagonalization, 
we do not analyze the existence of this phase within exact diagonalization. 
Instead, we analyze this phase using the Schwinger-boson mean field theory in the next section. 

\section{Schwinger Boson Mean-Field Theory} \label{sec:SchwingerMFT}

\let\vaccent=\v 
\renewcommand{\v}[1]{\ensuremath{\mathbf{#1}}} 
\newcommand{\gv}[1]{\ensuremath{\mbox{\boldmath$#1$}}}
\newcommand{\avg}[1]{\left< #1 \right>} 
\let\underdot=\d 
\renewcommand{\d}[2]{\frac{d #1}{d #2}} 
\newcommand{\pd}[2]{\frac{\partial #1}{\partial #2}}
\let\baraccent=\= 
\renewcommand{\=}[1]{\stackrel{#1}{=}} 
\newcommand{\bk}{\v{k}}
\newcommand{\bb}{\v{b}}
\newcommand{\vy}{\v{y}}
\newcommand{\vx}{\v{x}}
\newcommand{\vu}{\v{u}}
\newcommand{\vv}{\v{v}}
\newcommand{\vgm}{\gv{\gamma}}
\newcommand{\br}{\v{r}}
\newcommand{\HM}{\mathcal{H}}
\newcommand{\eref}[1]{(\ref{#1})}

In this section, we analyze the Heisenberg model introduced
in Eq.~\eqref{eq:H} using a Schwinger boson mean-field theory.\cite{Auerbach88,Sarker89,Sachdev91,Auerbach98}
The mean-field method can cover a wide range of circumstances where our
other methods may be limited. Unlike the strong-coupling expansion,
this method is not limited to small $J_x/J$ and $J_y/J$.
The incommensurate spiral state can be easily found by this mean-field 
theory, in contrast with exact diagonalization. Furthermore, the theory
can determine the splitting of the $x$- and $y$-spiral state degeneracy 
due to quantum fluctuations.
Finally, one can control the strength of quantum fluctuations
by varying the ``spin magnitude" $S$, connecting the $S \gg 1$ limit of
semi-classical ordering to the low-spin limit of $S=0.5$.
We consider the case $h=0$ for simplicity.

\subsection{Theory}

The Schwinger boson representation of the spin on site $i$ is
given by \cite{Auerbach98}
\begin{align}
	\v{S}_{i} = \frac{1}{2} \sum_{\alpha\beta} b^{\dag}_{i \alpha}
	\gv{\sigma}_{\alpha\beta}b_{i \beta},
\end{align}
Here, the local constraint of the form 
\begin{align}
	\sum_{\alpha} b^{\dag}_{i \alpha} b_{i \alpha} = 2S
\end{align}
is imposed 
so that the spin quantum number $S$ is given by the 
number of bosons per site.
The $\alpha,\beta$ spin labels run over $\uparrow,\downarrow$,
and $\sigma^a$ with $a=x,y,z$ are the Pauli matrices.

Heisenberg terms are quartic in the Schwinger bosons,
so we apply a mean-field decoupling to obtain a quadratic
Hamiltonian. The Heisenberg term is given by
\begin{align}
	\v{S}_{i} \cdot \v{S}_{j} = 
	\frac{1}{4}\sum_{a\mu\nu\rho\varepsilon}
	\sigma^{a}_{\mu\nu}\sigma^{a}_{\rho\varepsilon}
	b^{\dag}_{i \mu} b_{i \nu} 
	b^{\dag}_{j \rho} b_{j \varepsilon}.
\end{align}
To decouple this, we use the following Pauli matrix product identities
to decouple in an attractive channel for each possible sign of the
interaction:
\begin{align}
	\sum_{a}\sigma^{a}_{\mu\nu}\sigma^{a}_{\rho\varepsilon} = 
	\begin{cases}
		2 \delta_{\nu\rho}\delta_{\mu\varepsilon}
		- \delta_{\mu\nu}\delta_{\rho\varepsilon}
		\\ -2 \varepsilon_{\upsilon\rho}\varepsilon_{\nu\varepsilon}
		+ \delta_{\mu\nu}\delta_{\rho\varepsilon}
	\end{cases}.
\end{align}
Here, $\varepsilon_{\mu\nu}$ is the totally antisymmetric tensor
defined with $\varepsilon_{\uparrow\downarrow} = +1$.
This gives the relations for ferromagnetic and antiferromagnetic interactions,
\begin{align}
	-\v{S}_{i} \cdot \v{S}_{j} &= -\frac{1}{2}
	\hat{B}_{ij} \hat{B}^{\dag}_{ij}
	+ S(S+1),
	\\
	\v{S}_{i} \cdot \v{S}_{j} &= -\frac{1}{2}
	\hat{A}^{\dag}_{ij}\hat{A}_{ij}
	+ S^2,
	\label{eqn:SintermsofAB}
\end{align}
where we have defined the operators
\begin{align}
	\hat{B}_{ij} &= \sum_{\alpha} b^{\dag}_{i \alpha}b_{j\alpha},
	\\
	\hat{A}_{ij} &= \sum_{\alpha\beta} 
	b_{i\alpha}\varepsilon_{\alpha\beta}b_{j\beta}.
\end{align}

We enforce the boson number constraint by introducing 
a Lagrange multiplier term
\begin{align}
	\sum_{i}\lambda_{i} 
	\left( b^{\dag}_{i\alpha}b_{i\alpha} - 2S \right).
\end{align}
We treat this on average by taking $\lambda_{i}$ 
constant on each of the four sublattices.

We perform a mean-field decoupling of the terms given by \eref{eqn:SintermsofAB}.
For instance, 
\begin{align}
	\hat{A}^{\dag}_{ij} \hat{A}_{ij} \to 
	{A}^{*}_{ij} \hat{A}_{ij} + 
	\hat{A}^{\dag}_{ij} {A}_{ij} -
	|{A}_{ij}|^2  ,
\end{align}
with $A_{ij}=\langle \hat{A}_{ij}\rangle$. 
Assuming the translational invariance, there are eight $B$ mean fields, four for the nearest-neighbour
interactions in 
the $\hat{x}$ direction, four for the $\hat{y}$ direction, and two $A$ mean 
fields for the diagonal antiferromagnetic interactions.
Following this, we perform a Fourier transform defined by 
\begin{align}
	b_{\bk,X, \mu} = \frac1{\sqrt{N_{\rm uc}}} \sum_{\br\in A}e^{i\bk\cdot(\br+\delta\br_X)} 
	b_{\br,X, \mu}.
\end{align}
$X=A,B,C,D$ labels a position in a unit cell as in Fig.~\ref{fig:lattice}(b),
so we rewrite the label $i \to (\br,X)$. 
$\Nuc$ is the number of unit cells in the system. 
We have used the $A$ site position $\br$ to label a unit cell, and
$\delta\rvec_X$ to represent the position of each site relative to the $A$ site, 
\begin{equation}\label{eq:drX}
 \delta\rvec_A=0,~\delta\rvec_B=\hat{x},~
 \delta\rvec_C=\hat{y},~\delta\rvec_D=\hat{x}+\hat{y} .
\end{equation}

At this point, we assume the chemical equivalence of the
four sites in the unit cell, by taking $\lambda_{i} = \lambda$ the
same for all sites in the lattice.
In conjunction, we use an ansatz for the mean fields consistent with the
choice of chemical potential. Namely, we take the $\hat{x}$-direction $B$
to be equal $(B_x)$, and do the same for the $\hat{y}$-direction $B$ ($B_y$). 
The two $A$ mean fields are between sites $A$ and $D$ ($A_{AD})$, and $B$ and
$C$ ($A_{BC}$).
Furthermore, we take all mean fields to be real.
We take the two $A$ mean fields to have equal magnitude, so that 
$A_{AD}=A$ and $A_{BC} = \pm A$ gives two possible choices for the relative signs of A.
With these assumptions, the mean field theory can still describe all the relevant phases 
expected to appear in the model. 

We are left, in the case of a gapped dispersion,
with the mean-field Hamiltonian 
\begin{align}
	\HM_{MF} = H_C + \sum_{\bk} \v{b}^{\dag}_{\bk} \HM_{\bk} \v{b}_{\bk},
	\label{eqn:hamilmf}
\end{align}
where
\begin{align}
	H_C =& -2J_xB_x^2-2J_yB_y^2+JA^2-\lambda(8S+4),
	\nonumber \\
	\v{b}_{\bk}^T =& 
	\big(
	b_{\bk A\uparrow},
	b_{\bk B\uparrow},
	b_{\bk C\uparrow},
	b_{\bk D\uparrow},
	\nonumber \\ 
    &
	b_{-\bk A\downarrow}^{\dag},
	b_{-\bk B\downarrow}^{\dag},
	b_{-\bk C\downarrow}^{\dag},
	b_{-\bk D\downarrow}^{\dag}
	\big).
\end{align}
The matrix $\HM_{\bk}$
is given by
\begin{align}
	\HM_{\bk} = \begin{pmatrix} C_{\bk} & D_{\bk} \\ D_{\bk}^\dagger & C_{\bk} \end{pmatrix},
		\label{eqn:hamilk}
\end{align}
\begin{widetext}
\begin{align}
	C_k = \begin{pmatrix}
	\lambda & 
	J_xB_x\cos k_x
	& 
	J_yB_y\cos k_y
	& 0 \\
	J_xB_x\cos k_x &
	\lambda & 0 &
	J_yB_y\cos k_y
	\\
	J_yB_y\cos k_y
	& 0 &
	\lambda &
	J_xB_x\cos k_x \\
	0 & 
	J_yB_y\cos k_y & 
	J_xB_x\cos k_x &
	\lambda
\end{pmatrix},
\end{align}
\begin{align}
	D_k = \begin{pmatrix}
	0 & 0 & 0 &
	-\frac{J}{2}A_{AD} e^{-i(k_x+k_y)} \\
	0 & 0 & -\frac{J}{2}A_{BC} e^{-i(k_x-k_y)} 
	& 0 \\ 0 & 
	\frac{J}{2}A_{BC} e^{i(k_x-k_y)}
	& 0 & 0 \\
	\frac{J}{2}A_{AD} e^{i(k_x+k_y)}
	& 0 & 0 & 0 \\
\end{pmatrix},
\end{align}
\end{widetext}
with 
$k_x=\kvec\cdot\hat{x}$, 
$k_y=\kvec\cdot\hat{y}$, and $A_{AD},A_{BC}$ as defined above. 

The quadratic Hamiltonian in \eref{eqn:hamilmf} is diagonalized by
the Bogoliubov transformation $\bb_{\bk} = Z_k \vgm_{\bk}$,
with $\vgm$ defined in the same manner as $\bb$:
\begin{align}
	\vgm_{\bk}^T =& 
	\big(
	\gamma_{\bk A\uparrow},
	\gamma_{\bk B\uparrow},
	\gamma_{\bk C\uparrow},
	\gamma_{\bk D\uparrow},
	\nonumber  \\ 
    &
	\gamma_{-\bk A\downarrow}^{\dag},
	\gamma_{-\bk B\downarrow}^{\dag},
	\gamma_{-\bk C\downarrow}^{\dag},
	\gamma_{-\bk D\downarrow}^{\dag}
	\big).
\end{align}
This is a canonical transformation, where the operators in
$\vgm$ preserve the bosonic commutation relations of the operators 
in $\bb$.
These commutation relations are given by
\begin{align}
	[\vgm_\bk,\vgm^{\dag}_{\bk'}] = \delta_{\bk\bk'} \eta ,
\end{align}
where the $8\times8$ matrix $\eta$ can be written in $4\times4$
blocks:
\begin{align}
	\eta = \begin{pmatrix}
		I & 0 \\ 0 & -I
	\end{pmatrix}.
\end{align}
%
The proper commutation relation are obtained
by taking the columns of $Z_k$ to be the
eigenvectors $\vy_{m\bk}$ (with eigenvalues $\omega_{m\bk}$)
of $\eta\HM_{\bk}$.\cite{blaizotripka} 
Here, $m=1,\ldots,8$ labels the eight eigenvectors.

When the Hamiltonian $\HM_{\bk}$ is positive
definite, the corresponding spinon dispersions
$|\omega_{m\bk}|$ are gapped. 
This leads to a disordered ground state. 
In the case of a gapless dispersion, 
$\eta\HM_{\bk_0}\vy_{m'\bk_0} = 0$ for some $m'$,
so that both $\HM_{\bk_0}$ and 
$\eta\HM_{\bk_0}$ have a zero eigenvalue 
at the dispersion minimum $\bk_0$.
Condensation of such zero-energy bosons gives rise to a magnetically ordered state. 
To describe the condensation, 
we replace the operators
$\bb_{\bk_0}$ with macroscopic constant values,
\begin{equation}
\sum_{m'} \vx_{m'\bk_0} = \sum_{m'} c_{m'\bk_0}\sqrt{N_{\mathrm{uc}}}\vy_{m'\bk_0},
\end{equation}
where $\vy^{\dag}_{m'\bk_0}\vy_{m'\bk_0}=1$.
With the condensate contribution, the diagonalized Hamiltonian is
\begin{align}
	\HM_{MF} = H_C + \sum_{\bk} \sum_{m=1}^8 
	\vgm^{\dag}_{m\bk} |\omega_{m\bk}| \vgm_{m\bk}
	\nonumber \\
	+ \sum_{\bk_0} \sum_{m'}
	\vx^{\dag}_{m'\bk_0} \HM_{\bk_0} \vx_{m'\bk_0}.
\end{align}
Spin ordering is found at twice the spinon minimum wavevector, $2\bk_0$, 
governing relative spin orientation between unit cells.

Given the diagonalized Hamiltonian, we must solve for the mean-field
values of $A,B_x,B_y$ and $\lambda$, as well as any
condensate vectors $\vx_{m'\bk_0}$ and associated
minimum wavevectors $\bk_0$. It turns out that the solution depends only
on the total condensate density 
$\sum_{\bk_0}\sum_{m'} |c_{m'\bk_0}|^2$. 
We have the mean-field equations
\begin{align}
	\pd{\avg{\HM_{MF}}}{A} = 
	\pd{\avg{\HM_{MF}}}{B_x} = 
	\pd{\avg{\HM_{MF}}}{B_y}  
	\nonumber \\
	=\pd{\avg{\HM_{MF}}}{\lambda} = 
	\pd{\avg{\HM_{MF}}}{\vx_{m'\bk_0}} = 
	0,
\end{align}
which we solve self-consistently
for a given set of parameters $J_x/J$, $J_y/J$, and $S$,
and the associated values of $m'$ and $\bk_0$.

\subsection{Results}

\begin{figure}
	\includegraphics[scale=0.85]{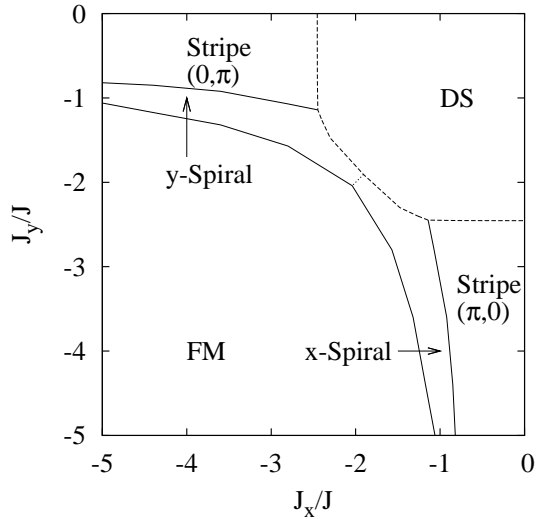}
	\caption{Schwinger boson phase diagram for $S=0.15$.
	Solid lines indicate second-order transitions, while dashed lines indicate first-order transitions. 
	The dotted line separates the two spiral phases. 
	The disordered dimer state (DS)
	expands significantly compared to the
	$S=0.5$ case in Fig.~\ref{fig:phase_SB}.
	The stripe state expands moderately,
	pushing the ferromagnetic (FM) state to moderately larger
	$J_x$ and $J_y$. In between these phases, the spiral state is
	found in a significantly reduced area. With an even smaller
	$S=0.1$, the spiral state will have disappeared entirely.
	}
	\label{fig:total0.15}
\end{figure}

In the semi-classical limit $S\to\infty$, we recover the classical phase
diagram shown earlier in Fig.~\ref{fig:phase_cl}.
We present the phase diagram as a function of $J_x/J$ and $J_y/J$ for the
cases of $S=0.5$ (Fig.~\ref{fig:phase_SB}) and 
$S=0.15$ (Fig.~\ref{fig:total0.15}). Since a mean-field theory
is expected to underestimate quantum fluctuations, 
it can be instructive to look at spin values
smaller than the actual case.
The Schwinger boson mean-field theory finds several magnetically ordered phase,
and one disordered phase.
The magnetically ordered phases are the ferromagnetic, spiral and stripe
phases seen in the classical limit, while the disordered 
phase features isolated dimers.
Below we summarize how each phase is described in the Schwinger boson formalism. 

\paragraph*{Ferromagnetic Phase.}
In this state, the antiferromagnetic order parameter $A=0$, while
both ferromagnetic order parameters take their maximum values, $B_x=B_y=2S$.
The spinon minimum wavevector $\bk_0 = \v{0}$, as is the spin ordering 
wavevector, so that the spins are fully polarized throughout 
the entire lattice.

\paragraph*{Stripe Phase.}
In this state, the ferromagnetic order parameter associated with the
smaller of $J_x$ and $J_y$ is zero. 
The ordering wavevector is $\v{0}$, and a stripe magnetization pattern
is found in the direction of the larger ferromagnetic interaction.
Depending on the relative signs of the $A$ parameters,
the spinon minimum wavevector is either $(0,\pi)$ or $\v{0}$, 
for $J_x>J_y$, both leading to the same ordered state of same energy.

\paragraph*{Spiral Phase.}
This is an incommensurate magnetically ordered state with a nonzero ordering
wavevector in the direction of smallest ferromagnetic interaction:
$\bk_0 = (k_x,0)$ for $J_y > J_x$. 
All of the mean-field parameters are nonzero.
There are two inequivalent spinon minimum wavevectors
$\pm \bk_0$ found in the spiral state. They both lead to the same type
of spiral order, described in Sec.~\ref{sec:classical}.
The relative sign of the two $A$ mean fields yields 
either the $x$-spiral or $y$-spiral ordering. 
We find that the classical degeneracy between $x$- and $y$-spiral ordering
is broken by quantum fluctuations, determining the ordering direction. 
For $J_x>J_y$, the $y$-spiral state
has a lower energy, while the $x$-spiral state has lower energy for $J_y>J_x$.

\paragraph*{Disordered Phase.}
The disordered phase is gapped, with no boson condensate, and has
no ferromagnetic correlations: $B_x=B_y=0$. 
The antiferromagnetic dimer order parameter $A\neq0$, leaving
a decoupled dimer state. This is a mean field description corresponding
to the dimer singlet state.

We consider the effect on the phase diagram of lowering spin from the
large-$S$ semiclassical limit, by comparing the 
classical, $S=0.5$, and $S=0.15$ phase diagrams 
shown in Figs.~\ref{fig:phase_cl}, \ref{fig:phase_SB} 
and \ref{fig:total0.15}, respectively.
As $S$ decreases, the 
disordered phase appears and expands around $J_x=J_y=0$, pushing
out the magnetically ordered phases to larger $J_x$ and $J_y$.
Of particular notes is 
the shift of the ferromagnetic phase boundary to larger $J_x$ and $J_y$  
in contradiction with the exact boundary for 
$J_x=J_y=J$ seen in Appendix \ref{app:exactGS}. 
Similarly, the stripe state boundary 
shifts 
to larger $J_y$ for 
$J_x>J_y$, and larger $J_x$ for $J_x<J_y$.
Between these aforementioned phase boundaries, 
we find the spiral state, which consequently gets pushed out to
larger $J_x$ and $J_y$. This state shrinks as $S$ decreases, having
almost disappeared at $S=0.15$, as seen in Fig.~\ref{fig:total0.15}.
In contrast with the case of the purely antiferromagnetic Shastry-Sutherland
lattice,\cite{Chung01}
we see no other disordered states than the dimer one down to $S=0.1$.
In particular, there are no short-range ordered analogues of the ferromagnetic, stripe,
or spiral ordered states. 

\section{Comparison with $\text{(CuCl)LaNb}_2\text{O}_7$} \label{sec:compound}
	
In this section we compare our theoretical results with the experimental data on (CuCl)LaNb$_2$O$_7$. 
The magnetization process of the compound shows a smooth increase 
between $B_{c1}=10.3$ T and $B_{c2}=30.1$ T,\cite{Kageyama05_mag} as in Fig.~\ref{fig:mag_h}(b).
Therefore, the compound should be located inside the DS-II regions in Fig.~\ref{fig:phase_ED} within the $J$-$J_x$-$J_y$ model.
Assuming the Land\'e factor $g=2$, 
the above magnetic fields are translated into energy units as
\begin{equation}\label{eq:hc1hc2}
\begin{split}
 h_{c1}=g\mu_B B_{c1}=1.2~ \text{meV},\\
 h_{c2}=g\mu_B B_{c2}=3.5~ \text{meV}. 
\end{split}
\end{equation}
According to a recent inelastic neutron scattering experiment,\cite{Tassel10} 
the triplet excitations at $B=0$ range from approximately 1.2 to 3.0 meV. 
Good agreement of the lower bound of the excitations with $h_{c1}$ suggests that 
the emergence of the magnetization is associated with the condensation of single-triplon excitations. 
Indeed, a Bose-Einstein condensation of some sort of bosonic excitations have been observed 
in the field range between $B_{c1}$ and $B_{c2}$.\cite{Kitada07} 
Here we do not address the possibility of a condensation of bound magnons proposed in Refs.~\onlinecite{Kageyama05_mag,Ueda07}, 
since it does not seem to occur within the $J$-$J_x$-$J_y$ model (except for the case $J_x,J_y>0$) as discussed in Sec.~\ref{subsec:SC2}.

Below we first discuss the experimental data within the $J$-$J_x$-$J_y$ model, 
using the second-order perturbative results of Sec.~\ref{subsec:SC2}. 
Then we discuss possible effects of other exchange couplings $J_{2b}$ and $J_4'$ shown in Fig.~\ref{fig:lattice}(a), 
within the first-order perturbation theory.

\subsection{$J$-$J_x$-$J_y$ model}\label{subsec:compound_JJxJy}

\begin{figure}
\begin{center}
\includegraphics[width=0.50\textwidth]{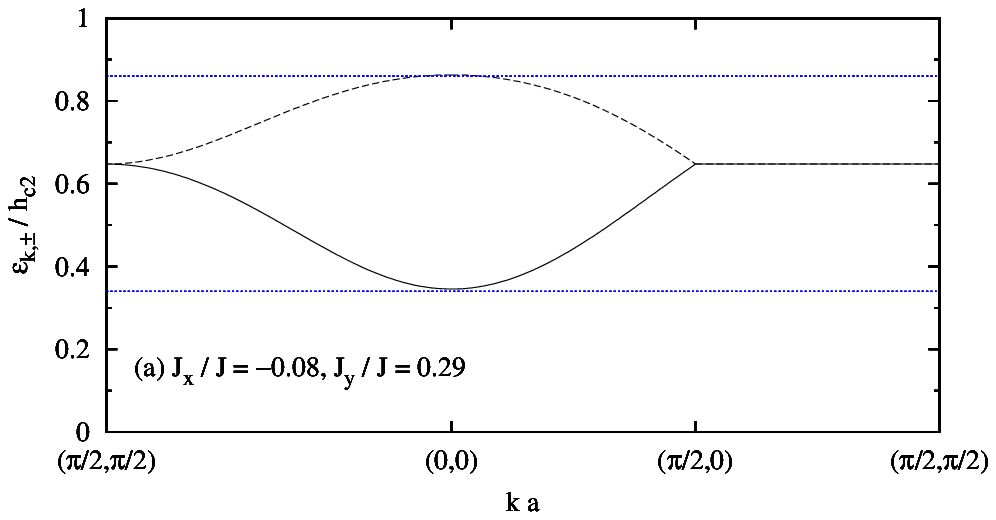}
\includegraphics[width=0.50\textwidth]{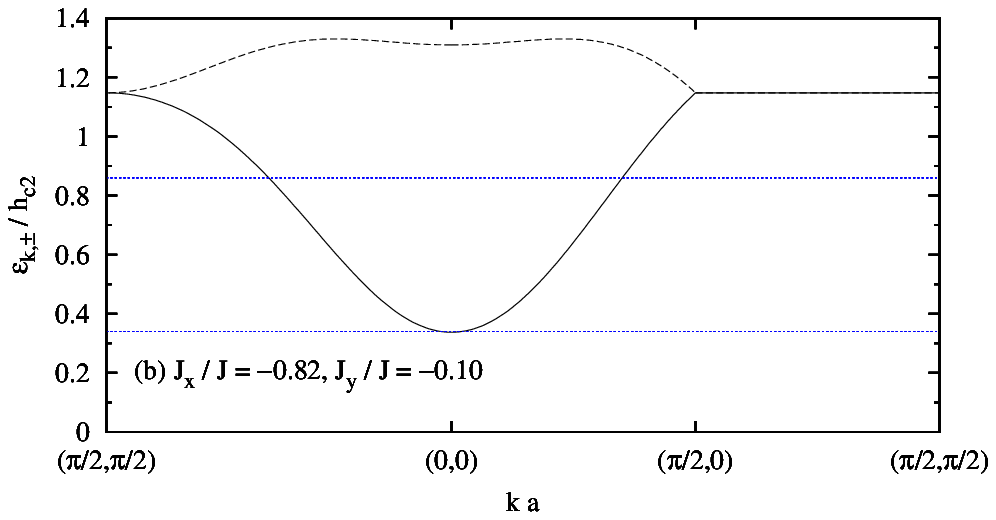}
\end{center}
\caption{(Color online)
Triplon dispersions \eqref{eq:bandpm_JJxJy} in the $J$-$J_x$-$J_y$ model 
in the cases of (a) ferromagnetic $J_x<0$ and antiferromagnetic $J_y>0$ in Eq.~\eqref{eq:estimate_JxJy}  
and (b) both ferromagnetic $J_x,J_y<0$ in Eq.~\eqref{eq:estimate_JxJy_FF}. 
Black solid and broken lines correspond to $\epsilon_{\bk,-}$ and $\epsilon_{\bk,+}$, respectively. 
These are plotted along a triangle connecting $(\pi/2,\pi/2)$, $(0,0)$, and $(\pi/2,0)$ in the Brillouin zone. 
Blue horizontal dotted lines indicate the energy range of triplet excitations (from $0.34 h_{c2}$ to $0.86 h_{c2}$) observed in Ref.~\onlinecite{Tassel10}. 
To compare with the magnetic layer of (CuCl)LaNb$_2$O$_7$ in Fig.~\ref{fig:lattice}(a), 
we have replaced $\bk\to \bk a$, where $a\approx 3.9 \AA$ is the distance between neighboring Cu$^{2+}$ ions. 
}
\label{fig:band_JJxJy}
\end{figure}

Using the above data of the magnetization process and the triplet excitations, 
the two ratios calculated in Fig.~\ref{fig:hc1} are given by 
\begin{equation}
 \Delta h/h_{c2}=0.66,~~
 \Delta\epsilon/h_{c2} = 0.52. 
\end{equation}
These can be used to fix the values of $(J_x/J,J_y/J)$, as shown by asterisks in Fig.~\ref{fig:hc1}. 
Adjusting $J$ to give $h_{c2}$ in Eq.~\eqref{eq:hc1hc2}, we obtain the estimates 
\begin{equation}\label{eq:estimate_JxJy}
 J= 2.2~\text{meV},~~
 J_x/J = -0.08,~~
 J_y/J = 0.29.
\end{equation}
Another parameter set where the values of $J_x$ and $J_y$ are interchanged is also possible. 
In these naive estimates within the $J$-$J_x$-$J_y$ model, 
one of $J_x$ and $J_y$ is ferromagnetic while the other is antiferromagnetic. 
This is in contrast to both ferromagnetic values $J_x,J_y<0$ predicted in the electronic structure calculation.\cite{Tassel10}

We note, however, that the estimation depends on the interpretation of the experimental data. 
In addition to the excitations ranging between $1.2$ and $3.0$ meV above, 
an earlier inelastic neutron scattering experiment\cite{Kageyama05} also detected 
excitations centered around $5.0$ meV ($\approx 1.43 h_{c2}$) with small scattering intensities. 
In Ref.~\onlinecite{Kageyama05}, these excitations are interpreted as bound states of triplets. 
If we regard that these also originate from triplet excitations, the range $\Delta \epsilon$ is much larger. 
Then in Fig.~\ref{fig:hc1}, the estimates of $(J_x/J,J_y/J)$ can change, keeping the relation $\Delta h/h_{c2}=0.66$, 
to the case of ferromagnetic $J_x,J_y<0$. 
For example, in the parameter set 
\begin{equation}\label{eq:estimate_JxJy_FF}
 J= 4.4~\text{meV},~~
 J_x/J = -0.82,~~
 J_y/J = -0.10, 
\end{equation}
the triplon excitation energy in Eq.~\eqref{eq:band_JJxJy} ranges up to $1.32h_{c2}$. 
It is expected that by increasing $|J_x|$ further, the energy upper bound would reach $1.43 h_{c2}$ observed in experiment.  
We do not discuss a larger-$|J_x|$ region,  
since our perturbative result would be less reliable there. 

To further discuss the consistency between the present model and (CuCl)LaNb$_2$O$_7$, 
it would be useful to look at the detailed shapes of the triplon dispersions.  
Here we plot the triplon dispersions 
using the perturbative result in Eq.~\eqref{eq:band_JJxJy}.  
When more detailed information of the triplet excitations is provided from experiments, 
these plots can be used to test the $J$-$J_x$-$J_y$ model and to determine the signs of $J_x$ and $J_y$. 
Equation~\eqref{eq:band_JJxJy} contained only a single band, 
since the single-triplon hopping problem had the same periodicity as the square lattice (of dimer centers). 
To discuss the physical excitations, however, 
one needs to take into account the fact that there are two dimers in a unit cell of the original lattice, 
and to fold the Brillouin zone in such a way as to identify $\bk$, $\bk+(\pi,0)$, and $\bk+(0,\pi)$. 
Consequently, the first Brillouin zone is given by a square ranging over $-\pi/2<k_{x,y}\le \pi/2$. 
After the folding, there are two dispersions expressed as
\begin{equation}\label{eq:bandpm_JJxJy}
 \epsilon_{\bm{k},\pm} = -\mu + 4t' \pm 4|t|\cos k_x\cos k_y - 16t' \cos^2 k_x \cos^2 k_y. 
\end{equation}
These dispersions are plotted in Fig.~\ref{fig:band_JJxJy} 
for the parameter sets in Eqs.~\eqref{eq:estimate_JxJy} and \eqref{eq:estimate_JxJy_FF}. 
While Fig.~\ref{fig:band_JJxJy}(a) shows a nearly symmetric shape centered around $0.66h_{c2}$, 
Fig.~\ref{fig:band_JJxJy}(b) shows a highly asymmetric shape. 


\subsection{$J$-$J_x$-$J_y$-$J_{2b}$-$J_4'$ model}\label{subsec:compound_JJxJyJ2bJ4p}
\newcommand{\Jt}{\tilde{J}}

\begin{figure}
\begin{center}
\includegraphics[width=0.50\textwidth]{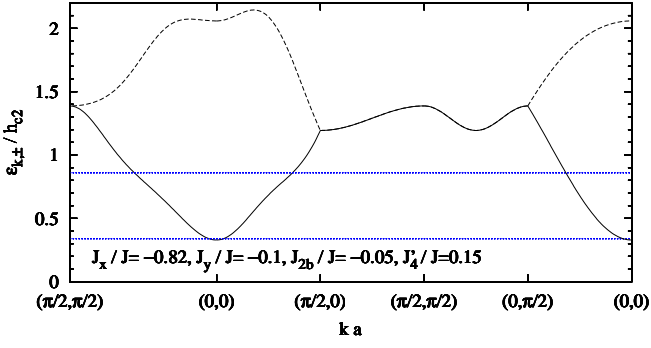}
\end{center}
\caption{(Color online)
Triplon dispersions \eqref{eq:band_JJxJyJ2bJ4p} in the $J$-$J_x$-$J_y$-$J_{2b}$-$J_4'$ model. 
Compared to Fig.~\ref{fig:band_JJxJy}, the path $(\pi/2,\pi/2) \to (0,\pi/2) \to (0,0)$ is added, 
since the triplon dispersions are anisotropic in the $k_x$ and $k_y$ directions in the presence of nonzero $J_{2b}$ or $J_4'$.  
}
\label{fig:band_ex1}
\end{figure}

In addition to $J$, $J_x$, and $J_y$, 
the model of Ref.~\onlinecite{Tassel10} also contains some other exchange couplings such as $J_{2b}$ and $J_4'$, shown in Fig.~\ref{fig:lattice}(a). 
Here we discuss their effects on the triplon dispersions. 

In the first-order perturbation theory, 
the additional couplings above introduce triplon hopping and interaction terms 
between the dimers they connect, similar to Eq.~\eqref{eq:Heff1_ori}. 
Using the effective Hamiltonian, triplon dispersions at $h=0$ are calculated as
\begin{equation}\label{eq:band_JJxJyJ2bJ4p}
 \epsilon_{\bk,\pm} = P_\bk \pm \sqrt{ Q_\bk^2 + R_\bk^2}, 
\end{equation}
with
\begin{subequations}
\begin{align}
 &P_\bk = J -\frac{J_4'}2 \cos 4k_x \cos 2k_y, \\
 &Q_\bk = -(J_x-J_y) \cos k_x \cos k_y + J_{2b} \cos 3k_x \cos k_y, \\
 &R_\bk = \frac{J_4'}2 \sin 4k_x \sin 2k_y. 
\end{align}
\end{subequations}
Around $\bk=\bm{0}$, these dispersions are expanded as
\begin{equation}
\begin{split}
 \epsilon_{\bk,\pm} =& J \pm |\Delta J| - \frac{J_4'}2 
 \mp \left[ \frac{|\Delta J|}2 + 4 ({\rm sgn}~\Delta J) J_{2b} \mp 4J_4' \right] k_x^2\\
 &\mp \left( \frac{|\Delta J|}2 \mp J_4' \right) k_y^2 + \Ocal (k^4),
\end{split}
\end{equation}
with $\Delta J:= -J_x+J_y+J_{2b}$. 
This dispersion implies that the minimum of the triplon excitation energy is given by 
\begin{equation}\label{eq:emin}
 \epsilon_{\bm{0},-} = J - |\Delta J| - \frac{J_4'}2, 
\end{equation}
if the following inequalities are satisfied:
\begin{gather}\label{eq:emin_condition}
 \frac{|\Delta J|}2 + 4 ({\rm sgn}~\Delta J) J_{2b} + 4J_4' >0,~~~
 \frac{|\Delta J|}2 + J_4' >0. 
\end{gather}
Equation~\eqref{eq:emin} can be used to determine the critical fields as\cite{comment_hc2}
\begin{align}
 &h_{c1} = \epsilon_{\bm{0},-} = J - |\Delta J| - \frac{J_4'}2,\\ 
 &h_{c2} = J+J_x+J_y+J_{2b}+\frac{J_4'}2 + |\Delta J|.
\end{align}


The electronic structure calculation of Ref.~\onlinecite{Tassel10} produced the estimates  
\begin{equation}\label{eq:J_Tassel}
\begin{split}
 &J_x/J=-0.39,~ J_y/J=-0.38,\\
 &J_{2b}/J=-0.14,~ J_4'/J=0.18. 
\end{split}
\end{equation}
These parameter values satisfy the condition \eqref{eq:emin_condition} 
and give $h_{c1}/J=0.78$ and $h_{c2}/J=0.31$. 
We find $h_{c1}>h_{c2}$, which means that the system is in fact inside the DS-I region, 
and the magnetization jumps as in Fig.~\ref{fig:mag_h}(a). 
We confirmed the occurrence of a jump by directly performing an exact diagonalization for the $J$-$J_x$-$J_y$-$J_{2b}$-$J_4'$ model 
with the values of Eq.~\eqref{eq:J_Tassel}. 
In order to be consistent with the smooth increase of the magnetization observed in experiment,\cite{Kageyama05_mag} 
the estimates in Eq.~\eqref{eq:J_Tassel} must be modified. 

Given the difficulty in uniquely determining the exchange couplings on the basis of the current experimental data, 
we here discuss possible effects of $J_{2b}$ and $J_4'$ on the triplon dispersions, 
which could be tested in future experiments. 
We add small values of $J_{2b}$ and $J_4'$ to the parameter set in Eq.~\eqref{eq:estimate_JxJy_FF}, 
and plot the triplon dispersions Eq.~\eqref{eq:band_JJxJyJ2bJ4p} in Fig.~\eqref{fig:band_ex1}. 
Compared to Fig.~\ref{fig:band_JJxJy}(b), the dispersions show oscillating behaviors 
coming from the long-range triplon hopping induced by $J_{2b}$ and $J_4'$. 
These oscillating behaviors can be used as fingerprints for the existence of these couplings. 

\section{Conclusions} \label{sec:discussion}

Motivated by recent experiments on (CuCl)LaNb$_2$O$_7$, 
we have studied a spin-$\frac12$ model \eqref{eq:H} of ferromagnetically coupled dimers on the distorted Shastry-Sutherland lattice. 
Using the three different approaches 
(the strong-coupling expansion, exact diagonalization, and Schwinger boson mean field theory), 
we have determined the ground state phase diagram of this model, 
as shown in Figs.~\ref{fig:phase_ED} and \ref{fig:phase_SB}. 
We have shown that in the dimer singlet phase appearing for $|J_x|,|J_y|\lesssim J$, 
the magnetization process depends crucially on the spatial anisotropy of $J_x$ and $J_y$. 
In the DS-I region with weak anisotropy, the magnetization shows a jump, as in Fig.~\ref{fig:mag_h}(a). 
In the DS-II regions with strong anisotropy, the magnetization smoothly increases 
after the spin gap closes at a certain magnetic field, as in Fig.~\ref{fig:mag_h}(b). 
When $|J_x|$ or $|J_y|\gtrsim J$, 
quantum phase transitions to various magnetically ordered phases 
(ferromagnetic, collinear stripe, and spiral) occur. 
These magnetic phases also appear in the classical limit of the model as shown in Fig.~\ref{fig:phase_cl}. 
Providing a notable difference from the classical case, 
the Schwinger boson analysis has demonstrated that 
quantum fluctuations split the classical degeneracy of two kinds of spirals ($x$- and $y$-spirals). 


We have compared our theoretical results 
with the existing experimental data on (CuCl)LaNb$_2$O$_7$. 
A smooth magnetization process observed in the compound\cite{Kageyama05_mag} 
suggests that, at least within the $J$-$J_x$-$J_y$ model, 
the compound should be located in the DS-II region, 
and that there should be substantial anisotropy in $J_x$ and $J_y$. 
Comparing the second-order perturbative results 
with the experimental data for the magnetization process\cite{Kageyama05_mag} and the triplon excitations,\cite{Tassel10} 
we found that one of $J_x$ and $J_y$ may be ferromagnetic while the other may be antiferromagnetic. 
This is in contrast to both the ferromagnetic estimates in the previous electronic structure calculation.\cite{Tassel10} 
However, we leave open the possibility of both ferromagnetic $J_x$ and $J_y$, 
given the possible existence of higher-energy triplet excitations. 
We plotted the triplon dispersions for different signs of $J_x$ and $J_y$. 
These plots can be used to test the $J$-$J_x$-$J_y$ model further and to determine the signs of $J_x$ and $J_y$,  
when more detailed information of the triplon excitations is provided from experiments. 
We have also discussed the effects of other exchange couplings $J_{2b}$ and $J_4'$ [shown in Fig.~\ref{fig:lattice}(a)] 
on the triplon dispersions.  


Our analysis of the distorted Shastry-Sutherland model may provide a useful starting point  
for understanding the wide range of physics in the layered copper oxyhalide family. 
A quantum phase transition from a singlet ground state to a collinear stripe state observed by chemical substitution\cite{Uemura09,Kitada09,Tsujimoto10} 
may be interpreted as the transition from the DS-II region with $J_x<J_y$ to the stripe state with a propagation vector $\bm{q}=(0,\pi)$ in Fig.~\ref{fig:phase_ED}. 
Here we have excluded the case of $J_x>J_y$, since the stripe state with $\bm{q}=(\pi,0)$ appearing in this region 
is in fact a N\'eel state with $\bm{q}=(\pi,\pi)$ in the original square lattice of the compound in Fig.~\ref{fig:lattice}(a).
The ion-exchange method used to synthesize the layered copper oxyhalide family 
has a remarkable flexibility of changing the constituent atoms in various ways. 
We expect that the jump of the magnetization in the DS-I region 
and a spiral order with a unique structure as in Fig.~\ref{fig:spiral_Qx} 
will also be observed in future experiments on related compounds.


\acknowledgements

We are grateful to Eric Lee, Seung-Hun Lee, and Tsutomu Momoi 
for stimulating discussions. 
This work was supported by the NSERC of Canada, 
the Canada Research Chair program, 
and the Canadian Institute for Advanced Research.
Numerical diagonalization calculations were performed 
using TITPACK ver.~2.\cite{titpack} 


\appendix

\section{Exact ground states in the isotropic case $J_x=J_y$}  \label{app:exactGS}


Here we discuss the exact ground states of the Hamiltonian \eqref{eq:H} 
in the isotropic case $J_x=J_y(\equiv J')$. 
The discussion of this section is based on the argument of Shastry and Sutherland\cite{Shastry81} 
and its extension to the case of ferromagnetic $J'$. 
We decompose the Hamiltonian $H$ into triangular parts 
and solve a 3-spin problem as in Fig.~\ref{fig:spiral_Qx}(a). 
The 3-spin Hamiltonian is given by 
\begin{equation}\label{eq:Htri}
 H_\triangle = J'(\Svec_1\cdot\Svec_2 + \Svec_2\cdot\Svec_3) + \frac{J}2 \Svec_1\cdot\Svec_3 -\frac{h}2 (S^z_1+S^z_3). 
\end{equation}
Note that here the Zeeman coupling with the magnetic field is introduced only for the first and third spins and not for the second spin.  
The Zeeman term for the second spin is 
included in the Hamiltonians of different triangles. 
In general, the decomposition of $H$ to triangular parts is not unique. 
However, definition of $H_\triangle$ used in Eq.~\eqref{eq:Htri}, introduced by Shastry and Sutherland,\cite{Shastry81}  
will be useful in the following discussion. 


When $h=0$, $H_\triangle$ is $SU(2)$-symmetric, 
and its eigenstates consist of a quadruplet $\ket{q_\mu} $ 
(with $\mu=\pm\frac12,\pm\frac32$ labeling $S^z_\mathrm{tot}$) and two doublets. 
Because $H_\triangle$ is also symmetric under the permutation of the sites $1$ and $3$, 
the doublets are classified into ones symmetric and antisymmetric 
with respect to this operation, $\ket{d_{\pm\frac12}}$ and $\ket{d'_{\pm\frac12}}$ respectively. 
When $h\ne 0$, the $SU(2)$ symmetry of $H_\triangle$ is reduced to $U(1)$.
Then $\ket{q_{\pm\frac12}}$ and $\ket{d_{\pm\frac12}}$ are mixed to form new eigenstates, 
$\ket{\qt_{\pm\frac12}}$ and $\ket{\dt_{\pm\frac12}}$. 
The eigenstates are summarized as follows: 
\begin{subequations}\label{eq:3spin_state}
\begin{align}
 &\ket{q_{+\frac32}} =
 \ket{\!\ua\ua\ua}, \\
 &\ket{\qt_{+\frac12}(\gamma_+)} = \frac1{\sqrt{2+\gamma_+^2}} 
 (\ket{\!\ua\ua\da} + \ket{\!\da\ua\ua} +\gamma_+ \ket{\!\ua\da\ua}), \\
 &\ket{\dt_{+\frac12}(\gamma_+)} = \frac1{\sqrt{4+2\gamma_+^2}} 
 (\gamma_+ \ket{\!\ua\ua\da} + \gamma_+ \ket{\!\da\ua\ua} -2 \ket{\!\ua\da\ua}),\\
 &\ket{d^\prime_{+\frac12}} = \frac1{\sqrt{2}}
 (\ket{\!\ua\ua\da} - \ket{\!\da\ua\ua}), 
\end{align}
\end{subequations}
and the other eigenstates with $\mu=-1/2$ or $-3/2$ 
are obtained by reversing all the spins and replacing $\gamma_+$ with $\gamma_-$. 
The parameters $\gamma_\pm$ are given by
\begin{equation}
 \gamma_\pm = -\frac12 \pm \frac{h}{2J'} + \frac1{2J'^2} \sqrt{9J'^2 \pm J'h+h^2}.
\end{equation}
The corresponding eigenenergies are given by 
\begin{subequations}
\begin{align}
 &E(q_{\pm\frac32}) = \frac{J}8 + \frac{J'}2 \pm \frac{h}2,\\
 &E(\qt_{\pm\frac12}) = \frac{J}8 + \frac{\gamma_\pm J'}2, \\
 &E(\dt_{\pm\frac12}) = \frac{J}8 - \frac{J'}{\gamma_\pm}, \\
 &E(d'_{\pm\frac12}) = -\frac{3J}8.
\end{align}
\end{subequations}
Crucially, the two states $\ket{d'_{\pm\frac12}}$ have the same eigenenergies 
and are written as product states of a singlet on the sites $1$ and $3$, and an isolated spin state on the site $2$. 
The exact dimer singlet state $\ket{\Psi_{\mathrm DS}}$ in Eq.~\eqref{eq:DS} can be written
in terms of $\ket{d'_{\pm\frac12}}$ on a given triangle. 
The state $\ket{\Psi_{\mathrm DS}}$ is therefore an eigenstate of the triangular Hamiltonian $H_\triangle$.
If $\ket{d'_{\pm\frac12}}$ are the ground states of the triangular Hamiltonian, the dimer singlet
state $\ket{\Psi_{\mathrm DS}}$ will also minimize the energy of the triangular Hamiltonian. 
Consequently, $\ket{\Psi_{\mathrm DS}}$ becomes an exact ground state of $H$, 
since it minimizes the local energy of every triangular Hamiltonian.
We note that the degeneracy of $\ket{d'_{\pm\frac12}}$ is not split by a magnetic field $h$, 
which comes from the specific form of the Zeeman coupling in Eq.~\eqref{eq:Htri}. 

\begin{figure}
\begin{center}
\includegraphics[width=0.45\textwidth]{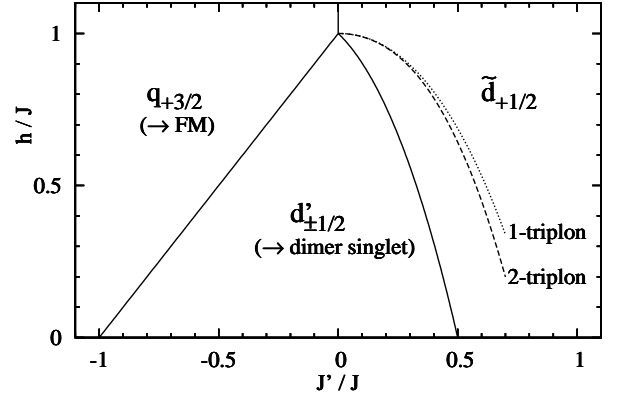}
\end{center}
\caption{
The ground state of the 3-spin Hamiltonian $H_\triangle$. 
The solid lines indicate the boundaries on which the 3-spin ground state changes. 
The exact ground states of $H$ derived from the 3-spin solution 
are indicated in the parentheses. 
For $J'>0$, the exact dimer singlet state $\ket{\Psi_{\mathrm DS}}$ continue to be the ground state of $H$ 
beyond the solid line (see the text). 
To indicate the range where $\ket{\Psi_{\mathrm DS}}$ actually remains the ground state, 
we show broken and dotted line. 
These lines respectively indicate the magnetic fields on which two- and one-triplon excitations condense 
according to the perturbative calculation of Ref.~\onlinecite{Momoi00}. 
}
\label{fig:phase_exact}
\end{figure}


Assuming $J>0$ and $h\ge 0$, 
the ground state of $H_\triangle$ is summarized in Fig.~\ref{fig:phase_exact}. 
For $h=0$, $\ket{d'_{\pm\frac12}}$ are the ground states of $H_\triangle$ when 
\begin{equation}
 -1< \frac{J'}J < \frac12. 
\end{equation}
These states remain the ground states of $H_\triangle$ 
up to a certain magnetic field, indicated by solid lines in Fig.~\ref{fig:phase_exact}. 
In this regime, the dimer singlet state $\ket{\Psi_{\mathrm DS}}$ is an exact ground state of $H$ 
with eigenenergy $-\frac38 N_s J$. 

For $J'<0$, the ground state of $H_\triangle$ changes to $\ket{q_{+\frac32}}$ 
as we pass the solid line in Fig.~\ref{fig:phase_exact}. 
In this regime, the ground state of $H$ is given by a fully polarized ferromagnetic state 
since it contains the state $\ket{q_{+\frac32}}$ locally on every triangle. 
Combined with the above result, the present exact argument can complete the phase diagram of $H$ in the case of ferromagnetic $J'$. 
Specifically, the phase boundary is given by $h=J+J'$. 

For $J'>0$, the ground state of $H_\triangle$ is replaced by $\ket{\tilde{d}_{+\frac12}}$ 
beyond the solid line in Fig.~\ref{fig:phase_exact}. 
The solid line is give by 
\begin{equation}\label{eq:h_ex}
 h = \frac{4(J+J')(J-2J')}{4J-J'}. 
\end{equation}
Outside of this region, we cannot use the 3-spin solution to predict the ground state of $H$, 
since $\ket{\tilde{d}_{+\frac12}}$ is not factorizable. 
However, one can show that $\ket{\Psi_{\mathrm DS}}$ is always an exact eigenstate of $H$, regardless of the values of $J$, $J'$ and $h$. 
Therefore, beyond the line \eqref{eq:h_ex}, $\ket{\Psi_{\mathrm DS}}$ should still remain the exact ground state of $H$ 
until it encounters a level crossing with another eigenstate. 
For $h=0$, 
this crossing has been found to occur at $J'/J\approx 0.69$ in an exact diagonalization study of finite clusters.\cite{Miyahara99} 
For $h>0$, the level crossing is associated with a condensation of triplon excitations. 
As shown in Ref.~\onlinecite{Momoi00}, a bound state of two triplons condense before one-triplon state does. 
In Fig.~\ref{fig:phase_exact}, we draw the two lines, derived from the strong coupling expansion to third order in Ref.~\onlinecite{Momoi00},   
where one- and two-triplon condensations occur.


\section{Hartree variational state} \label{app:Hartree}


Here we discuss an alternative derivation of the phase diagram in Fig.~\ref{fig:phase_SC1}, 
using the Hartree variational state.\cite{Momoi00,Penc07}  
The discussion goes in parallel with Sec.~\ref{sec:SC}, 
and some notations are common with this section. 
The Hartree variational state is given by a product of local states on dimers:
\begin{equation}\label{eq:Hartree}
 \ket{\Psi} = \prod_\Rvec \left( \cos\frac{\theta_\Rvec}2 \ket{t_{+1}}_\Rvec + e^{i\phi_\Rvec} \sin\frac{\theta_\Rvec}2 \ket{s}_\Rvec \right).
\end{equation}
This state smoothly connects between the dimer singlet state ($\theta_\Rvec=\pi$) at $h=0$ and the fully polarized state ($\theta_\Rvec=0$) at a high field. 
For $0<\theta_\Rvec<\pi$, this state describes a superfluid state of triplons, 
which breaks the $U(1)$ symmetry. 
We minimize the energy $\bra{\Psi}H\ket{\Psi}$ with respect to the variational parameters $\{\theta_\Rvec;\phi_\Rvec\}$ 
to find a variational ground state. 
Using the obtained values of $\{\theta_\Rvec;\phi_\Rvec\}$, one can determine the magnetic structure via the relations
\begin{subequations}\label{eq:Hartree_S}
\begin{align}
 &\langle S^x_{\Rvec1}+iS^y_{\Rvec1} \rangle = -\langle S^x_{\Rvec2}+iS^y_{\Rvec2} \rangle = - \frac{1}{2\sqrt{2}} e^{i\phi_\Rvec} \sin\theta_\Rvec,\\
 &\langle S^z_{\Rvec1} \rangle = \langle S^z_{\Rvec2} \rangle = \frac14 (1+\cos\theta_\Rvec).
\end{align}
\end{subequations}

Since $\ket{\Psi}\in V_0$, we find
\begin{equation}
 \bra{\Psi}H\ket{\Psi}=\bra{\Psi}P H P\ket{\Psi}=E_0 + \bra{\Psi}H^\eff_1\ket{\Psi}, 
\end{equation} 
where $P$ is the projection operator onto $V_0$. 
Since Eq.~\eqref{eq:Hartree} can also be viewed as a coherent state of $s_\Rvec$ spins, 
minimizing $\bra{\Psi}H^\eff_1\ket{\Psi}$ is precisely equivalent to finding the classical ground state of the XXZ model $H^\eff_1$. 
The solution to the classical problem\cite{Neel36,Yosida51,Kohno97} is essentially similar to the quantum solution 
(except for the detailed shape of the magnetization process), 
and leads to the same phase diagram of the original model $H$ as in Fig.~\ref{fig:phase_SC1}. 
In the entire parameter space, the obtained variational ground state has the same periodicity as the original model; 
the values of $\theta_R$ and $\phi_R$ are uniform in each subset of dimers ($\Dcal_1$ or $\Dcal_2$). 
Therefore, the ground states are characterized by four parameters $\theta_{1,2}$ and $\phi_{1,2}$ defined for $\Dcal_{1,2}$. 
In the DS-II region with $J_x<J_y$, the classical ground state of $H^\eff_1$ is given by a canted antiferromagnetic state with 
$\phi_1=\phi_2+\pi$ and $\theta_1=\theta_2$. 
Using Eq.~\eqref{eq:Hartree_S}, 
this leads, in the original model, to a canted stripe state with a propagation vector $(0,\pi)$ in the transverse component. 
Similarly, in the DS-II region with $J_x>J_y$, the classical solution of $H^\eff_1$ is given by $\phi_1=\phi_2$ and $\theta_1=\theta_2$, 
leading to a canted stripe state with a propagation vector $(\pi,0)$. 
The magnetization curve in the DS-II region is determined as in Fig.~\ref{fig:mag_h}(b), 
but the curves are all straight in the current variational approach. 
As we increase $|J_x|$ or $|J_y|$, $h_{c1}$ goes to zero, signaling a phase transition to a collinear stripe phase. 
We note that even after the transition, the variational ground state at $h=0$ remains the dimer singlet state $\ket{\Psi_{\rm DS}}$ with $\theta_1=\theta_2=\pi$. 
This is an artifact of the ansatz. 
In the DS-III region, the variational solution for $M=M_s/2$ is given by $(\theta_1,\theta_2)=(0,\pi)$ and $(\pi,0)$, 
indicating th existence of the plateau. 

\section{Single magnon from the polarized state}  \label{app:saturation}
\newcommand{\Jcal}{{\cal J}}

Here we consider a single-magnon excitation from the fully polarized state 
$\ket{\rm FM}=\ket{\ua\ua\ua\dots}$ 
and determine the saturation field $h_{c2}$. 
We introduce the Fourier-transformed basis
\begin{equation}\label{eq:Xk}
 \ket{X,\kvec} = \frac{1}{\sqrt{\Nuc}} \sum_{\rvec \in A} e^{i\kvec\cdot(\rvec+\delta\rvec_X)} S^-_{\rvec, X} \ket{\rm FM}, 
\end{equation}
where $X=A,B,C,D$ labels a position in a unit cell as in Fig.~\ref{fig:lattice}(b). 
We have used the $A$ site position $\rvec$ to label a unit cell, and
$\delta\rvec_X$ represents the position of each site relative to the $A$ site, as in Eq.~\eqref{eq:drX}. 
Using the basis \eqref{eq:Xk}, we define a $4\times4$ matrix $\Mcal (\kvec)$ by
\begin{equation}
 \Mcal_{XX'}(\kvec) = \bra{X,\kvec} H \ket{X',\kvec} - \delta_{XX'} \bra{{\rm FM}}H\ket{{\rm FM}}. 
\end{equation}
Here, the energy of the fully polarized state is subtracted. 
To represent the matrix $\Mcal (\kvec)$ in a compact form, 
we identify $X=A,B,C,D$ with fictitious two-spin-$\frac12$ states $\ua\ua,\ua\da,\da\ua,\da\da$ 
and introduce two sets of Pauli matrices, $\bm{\sigma}_1$ and $\bm{\sigma}_2$.
The obtained expression is 
\begin{equation} \label{eq:Mmat}
 \Mcal(\kvec) = \Jcal(\kvec) + \left( h-\frac{J}2-J_x-J_y \right) I,
\end{equation}
where
\begin{equation}
\begin{split}
 \Jcal (\kvec) &= J_x \sigma_2^x \cos k_x + J_y \sigma_1^x \cos k_y \\
 &+ \frac{J}2 (\sigma_1^x \cos k_x + \sigma_1^y \sin k_x) (\sigma_2^x \cos k_y + \sigma_2^y \sin k_y). 
\end{split}
\end{equation}
with $k_x=\kvec\cdot\hat{x}$ and $k_y=\kvec\cdot\hat{y}$. 
By diagonalizing $\Jcal (\kvec)$ and obtaining the eigenvalues $\lambda_\nu(\kvec)~(\nu=1,2,3,4)$, 
one obtains four magnon dispersions $\tilde{\lambda}_\nu(\kvec):=\lambda_\nu(\kvec)+ h-J/2-J_x-J_y$. 
Assuming that the leading instability from the fully polarized state is a condensation of single-magnon excitations, 
the saturation field $h_{c2}$ is found when the minimum of the lowest dispersion $\tilde{\lambda}_1(\kvec)$ touches zero. 
This leads to an expression 
$h_{c2} = \frac{J}2 + J_x+J_y+\lambda_{\rm min}$, 
where $\lambda_{\rm min}$ is the minimum of $\lambda_1(\kvec)$ over $\kvec$. 

We now determine $\lambda_{\rm min}$. 
One can show, through the formulation of Luttinger and Tisza,\cite{Luttinger46} 
that this problem is precisely equivalent to the classical ground-state problem. 
Therefore, the analysis can be done in parallel with Sec.~\ref{sec:classical}. 
In the collinear stripe phases of Fig.~\ref{fig:phase_cl}, 
the minimum of the dispersion is found at $\kvec=(0,0)$. 
At this point, $\Jcal(\bm{0})$ is expressed using only mutually commuting matrices, 
$\sigma_1^x$ and $\sigma_2^x$, and thus is diagonalized easily. 
The saturation field is 
\begin{equation}\label{eq:hc2_stripe}
 h_{c2} = J + J_x + J_y + |J_x-J_y|. 
\end{equation}
In the classically spiral regime, the minima of the dispersion are found at $\kvec=(Q_x,0)$ and $(0,Q_y)$, 
where $Q_x$ and $Q_y$ are given in Eq.~\eqref{eq:QxQy}. 
When one of $k_x$ and $k_y$ is set to zero, 
the diagonalization of $\Jcal(\kvec)$ is simple, because 
one of $\sigma_1^x$ and $\sigma_2^x$ becomes a good quantum number 
leaving a single-spin problem in a magnetic field. 
The saturation field is calculated as
\begin{equation}\label{eq:hc2_spiral}
 h_{c2} = \frac{JJ_xJ_y}4 \left( \frac2J + \frac1{J_x} + \frac1{J_y} \right)^2.
\end{equation}
We comment that in the isotropic case $J_x=J_y$, 
the minimum of the dispersion is obtained along a circle in $\kvec$ space, 
leading to an interesting pseudo-one-dimensional density of states of magnons.\cite{Momoi00} 

In closing, we stress that our argument for determining $h_{c2}$ in this section 
is based on the assumption that the leading instability from the fully polarized state 
is a condensation of single-magnon excitations. 
The present argument therefore does not apply to the DS-I region, 
where a direct transition occurs from the fully polarized state to a singlet state as one lowers the magnetic field.



\begin{thebibliography}{99}
\bibitem{Comment_PS}
A phase-separated state consist of fully occupied ($n_R=1$) and empty ($n_R=0$) domains of hard-core bosons.  
If the total particle number is increased gradually, 
one observes gradual expansion and shrink of the occupied and empty domains, respectively. 
If the chemical potential is changed, 
one instead observes a first-order transition between fully empty and occupied state which are homogeneous in the entire system. 
This leads to a jump in the magnetization in the spin Hamiltonian \eqref{eq:H}.
\bibitem{comment_FM}
The classical ferromagnetic phase boundary corresponds, in the quantum case, 
to the line where one-magnon excitations from the fully polarized state condense (namely, $h_{c2}$ in Eq.~\eqref{eq:hc2_spiral} becomes zero). 
In the transition from the ferromagnetic phase to the DS-I region in the spin-$\frac12$ case, 
the ground state directly change from the fully polarized state to a singlet state before the one-magnon condensation occurs. 
A good agreement of the quantum and classical phase boundaries means that 
the $E$-$M$ curve as in Fig.~\ref{fig:ene_M}(a) becomes almost straight around the phase boundary.
\bibitem{comment_hc2}
In principle, the minimum of the dispersion at $h=0$ in Eq.~\eqref{eq:emin} can be used only for determining $h_{c1}$. 
To calculate $h_{c2}$, one needs to consider a single-magnon excitation from the fully polarized state 
and to find the minimum of the dispersion. 
Within the first-order perturbation theory, however, the two problems can be solved in parallel, 
since the effective model is equivalent to an XXZ model and has the particle-hole symmetry.
\end{thebibliography}
\end{document}